\documentclass{emulateapj}

\usepackage{graphicx,times}
\usepackage{epstopdf, epsfig}
\usepackage{multirow}
\usepackage{amsmath,amssymb}
\usepackage{rotating}

\usepackage{CJKutf8}

\begin{document}

\title{The Hawaii SCUBA-2 Lensing Cluster Survey: Radio-detected Submillimeter Galaxies in the HST Frontier Fields}

\shortauthors{Hsu et al.}

\author{Li-Yen Hsu \begin{CJK*}{UTF8}{bsmi}(徐立研)\end{CJK*}\altaffilmark{1}, Vandana Desai\altaffilmark{2}, Eric J. Murphy\altaffilmark{2,3}, Lennox L. Cowie\altaffilmark{1}, Ian Heywood\altaffilmark{4}, Emmanuel Momjian\altaffilmark{5},\\Amy J. Barger\altaffilmark{1,6,7}, and Ian Smail\altaffilmark{8,9}}

\altaffiltext{1}{Institute of Astronomy, University of Hawaii, 2680 Woodlawn Drive, Honolulu, HI 96822, USA}
\altaffiltext{2}{Spitzer Science Center, California Institute of Technology, 1200 E. California Boulevard, Pasadena, CA 91125, USA}
\altaffiltext{3}{National Radio Astronomy Observatory, 520 Edgemont Road, Charlottesville, VA, USA}
\altaffiltext{4}{CSIRO Astronomy and Space Science, Australia Telescope National Facility, P.O. Box 76, Epping, NSW 1710, Australia}
\altaffiltext{5}{National Radio Astronomy Observatory, 1003 Lopezville Road, Socorro, NM 87801, USA}
\altaffiltext{6}{Department of Astronomy, University of Wisconsin-Madison, 475 North Charter Street, Madison, WI 53706, USA}
\altaffiltext{7}{Department of Physics and Astronomy, University of Hawaii, 2505 Correa Road, Honolulu, HI 96822, USA}
\altaffiltext{8}{Centre for Extragalactic Astronomy, Department of Physics, Durham University, South Road, Durham DH1 3LE, UK}
\altaffiltext{9}{Institute for Computational Cosmology, Durham University, South Road, Durham DH1 3LE, UK}

\begin{abstract}

In this second paper of the Hawaii SCUBA-2 Lensing Cluster Survey series, we cross-match SCUBA-2 maps with 3 and 6 GHz images from the Janksy-VLA Frontier Fields Legacy Survey for three 
cluster fields, MACS\,J0416.1--2403, MACS\,J0717.5+3745, and MACS\,J1149.5+2223. Within the {\it HST} coverage, 14 out of 44 850 $\mu$m sources have 3 GHz counterparts, five of which are 
also detected at 6 GHz. The 850 $\mu$m flux densities of these detected sources span from 0.7 to 4.4 mJy after correcting for lensing amplification. The median redshift of the sample is $z = 1.28^{+0.07}_{-0.09}$, much 
lower than the typical redshifts ($z = 2-3$) of brighter submillimeter galaxies in the literature. In addition, we find that our sources have lower dust temperatures than those of the brighter submillimeter galaxies. This 
is also confirmed by an analysis of the ratio between infrared star formation rate and 850 $\mu$m flux density. However, these 14 sources may not represent the general submillimeter population at the 
same flux range, given that the SCUBA-2 sources without radio counterparts are likely at higher redshifts. Detection of these sources would require deeper radio images or submillimeter interferometry.

\end{abstract}

\subjectheadings{cosmology: observations|  galaxies: formation  |  galaxies: starburst  |  gravitational lensing: strong | submillimeter: galaxies }

\section{Introduction}

Submillimeter galaxies (SMGs; reviews by \citealt{Blain2002Submillimeter-g,Casey2014Dusty-Star-Form}) are some of the most massively star-forming galaxies in the universe. They were first detected in deep-field 
maps (e.g., \citealt{Smail1997A-Deep-Sub-mill,Barger1998Submillimetre-w,Hughes1998High-redshift-s}) made with the Submillimeter Common-User Bolometer Array (SCUBA;  \citealt{Holland1999SCUBA:-a-common}) on the 15m James Clerk Maxwell Telescope (JCMT). Because SMGs are dusty and have high extinction, many of them are not detected in UV/optical surveys (e.g., \citealt{Barger2012Precise-Identif,Simpson2014An-ALMA-Survey-}). To 
understand how the most massive galaxies formed and how the star formation rate (SFR) density evolves with cosmic time, it is crucial to study these SMGs since 
they contribute a significant fraction ($>$ 10\%) of the star formation at high redshifts (e.g., \citealt{Barger2000Mapping-the-Evo,Barger2012Precise-Identif,Barger2014Is-There-a-Maxi,Chapman2005A-Redshift-Surv,Wang2006A-Near-Infrared,Serjeant2008The-SCUBA-Half-,2011MNRAS.415.1479W,Casey2013Characterisatio,Cowie2017A-Submillimeter}).

The SCUBA-2 camera \citep{2013MNRAS.430.2513H} on the JCMT is currently the most powerful instrument to carry out deep and wide-field surveys to search for SMGs. It covers 16 times the area of the previous SCUBA camera and has the fastest mapping speed at 450 and 850 $\mu$m among single-dish far-infrared (FIR) telescopes. However, the confusion limit \citep{Condon1974Confusion-and-F} of JCMT ($\sim$ 2 mJy at 850 $\mu$m) prevents 
the detection of fainter galaxies with infrared (IR) luminosities $< 10^{12} L_{\odot}$. As a result, there is little information about lower luminosity galaxies, which may be expected to have SFRs comparable to those of 
the UV/optical populations. Imaging of massive galaxy cluster fields is one way to reach fainter detection limits because background sources are gravitationally magnified. Previous studies have constructed number counts (e.g., \citealt{Smail1997A-Deep-Sub-mill,Smail2002The-nature-of-f,Cowie2002Faint-Submillim,Knudsen2008Probing-the-sub,Johansson2011A-LABOCA-survey,Chen2013Faint-Submillim,Chen2013Resolving-the-C,Fujimoto2016ALMA-Census-of-,Hsu2016The-Hawaii-SCUB}) or detected individual sources \citep{Watson2015A-dusty-normal-,Gonzalez-Lopez2017The-ALMA-Fronti} using submillimeter/millimeter observations of cluster fields.

To construct a large sample of faint SMGs that contribute the majority of extragalactic background light (EBL), we have been undertaking a SCUBA-2 program, the Hawaii SCUBA-2 Lensing Cluster Survey 
(Hawaii-S2LCS). This program maps nine massive clusters, including the northern five clusters in the {\it HST} Frontier Fields program \citep{Coe2015Frontier-Fields}. \cite{Hsu2016The-Hawaii-SCUB} present 
deep number counts at 450 and 850 $\mu$m based on SCUBA-2 observations of six cluster fields and three blank fields.

Given the low spatial resolution of single-dish telescopes (FWHM $\sim$ 7$\farcs$5 at 450 $\mu$m and $\sim$ 14$\farcs$5 at 850 $\mu$m for JCMT), interferometric follow-up is required to identify the 
multi-wavelength counterparts to submillimeter sources. While submillimeter interferometry is the most reliable way to do this, it is observationally expensive. Radio interferometry is an effective 
alternative that relies on the observed correlation between FIR emission and radio emission from local starburst galaxies \citep{Helou1985Thermal-infrare,Condon1992Radio-emission-}. Although the physics of this FIR-radio correlation is unclear, the non-thermal synchrotron emission from supernova remnants traces the dust-obscured star formation (e.g., \citealt{Murphy2009The-Far-Infrare,Ivison2010BLAST:-the-far-,Ivison2010The-far-infrare,Momjian2010High-sensitivit}). The disadvantage of radio identification is that it does not benefit from a negative $K$-correction, making it difficult to detect SMGs at $z >$ 3.

The Janksy-VLA Frontier Fields Legacy Survey (PI: Eric Murphy) aims to characterize the dust-obscured properties of high-redshift galaxies through Karl G. Jansky Very Large Array (VLA) imaging of all five 
{\it HST} Frontier Fields observable with the VLA at 3 and 6 GHz.  The goal rms sensitivity of these images is $\lesssim 1\,\mu$Jy on the image plane, which is achieved for all fields with data in hand. At 6 GHz, these data reach an angular resolution of 0$\farcs$3, similar to the resolution of {\it HST}/WFC3. These data will allow a variety of extragalactic studies, including radio morphologies of star-forming galaxies, obscured star formation out to $z \sim$ 8, the evolution of 
supermassive black holes, and the rapid evolution of galaxies in the lensing clusters themselves. Observations and data reductions of this survey are still ongoing, and the catalogs of detected sources will 
be presented in a forthcoming paper (Heywood et al. 2017, in preparation)

In this second paper of the Hawaii-S2LCS series, we present a sample of 14 SCUBA-2 850 $\mu$m sources identified with the Janksy-VLA Frontier Fields Legacy Survey in the fields of MACS\,J0416.1--2403, MACS\,J0717.5+3745, and MACS\,J1149.5+2223 (hereafter, MACSJ0416, MACSJ0717, and MACSJ1149). The details of the observations and data reduction are described in Section~2. Section~3 describes 
source extraction and sample selection. The derived properties of our sample are discussed in Section~4. In Section~5, we discuss the detectability of submillimeter sources in radio surveys and 
optical-near-infrared color selections. Section~6 summarizes our results. Throughout this paper, we assume a \cite{Kroupa2001On-the-variatio} initial mass function (IMF) and the 
concordance $\Lambda$CDM cosmology with $H_0=70~\rm km~s^{-1}~Mpc^{-1}$, $\Omega_M=0.3$, and $\Omega_\Lambda=0.7$.

 \section{Data}\label{sec:data}

\subsection{SCUBA-2 Images}

We combined all of our SCUBA-2 data taken between February 2012 and March 2016. We used the CV DAISY scan pattern to detect sources out to $\sim$ 6$'$ from the cluster centers. Most of our observations 
were carried out under band 1 (the driest weather; $\tau_{\rm 225GHz} < 0.05$) conditions, but there are also data taken under band 2 ($0.05 < \tau_{\rm 225GHz} < 0.08$) or good band 3 conditions 
($0.08 < \tau_{\rm 225GHz} < 0.1$). We summarize the details of these observations in Table~\ref{tab:table1}.

\begin{figure*}
\includegraphics[width=6cm]{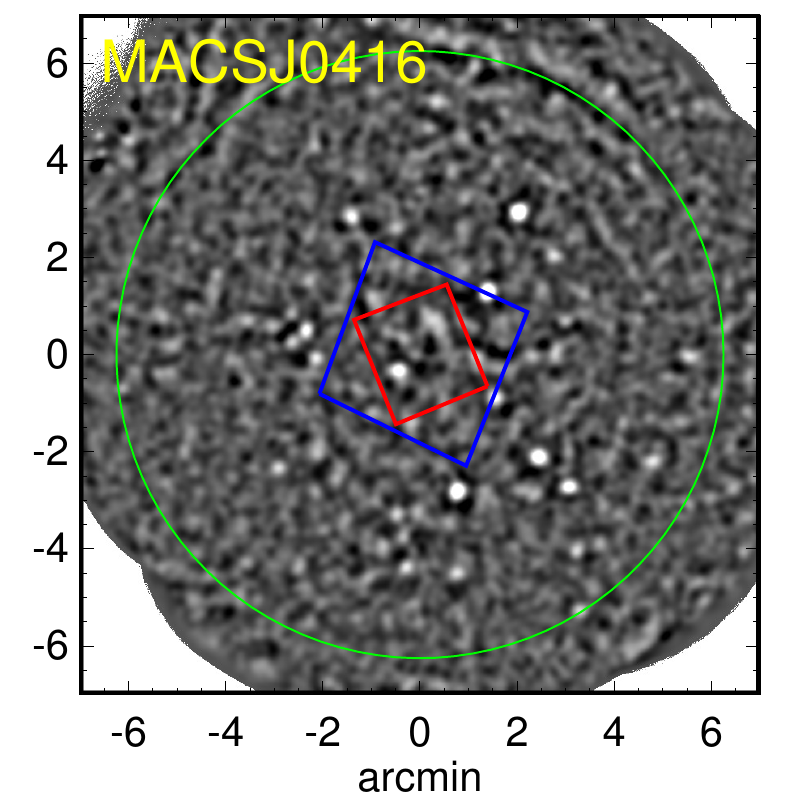}
\includegraphics[width=6cm]{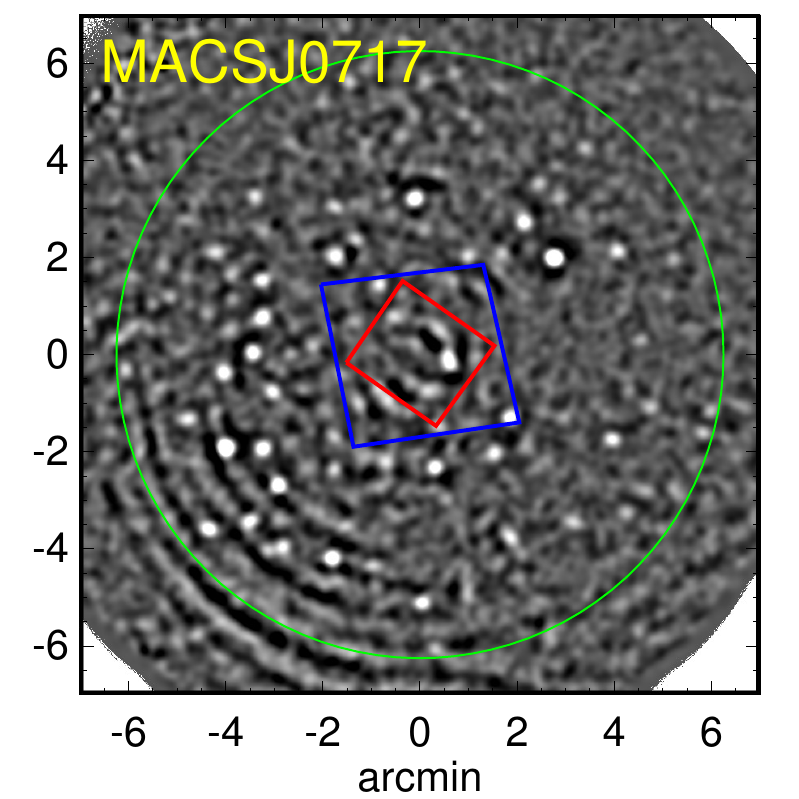} 
\includegraphics[width=6cm]{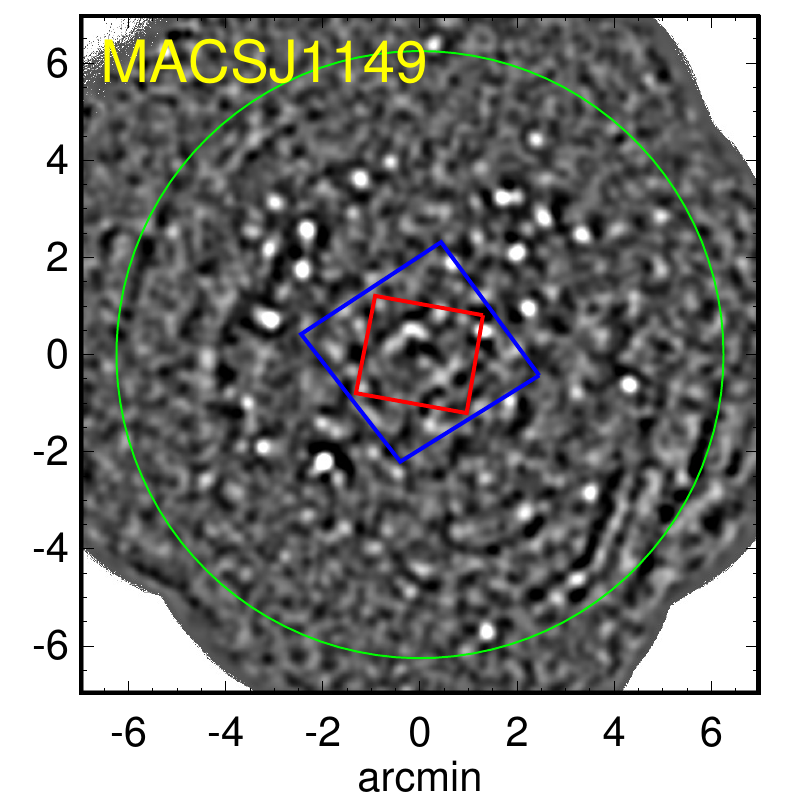}\\
\includegraphics[width=6cm]{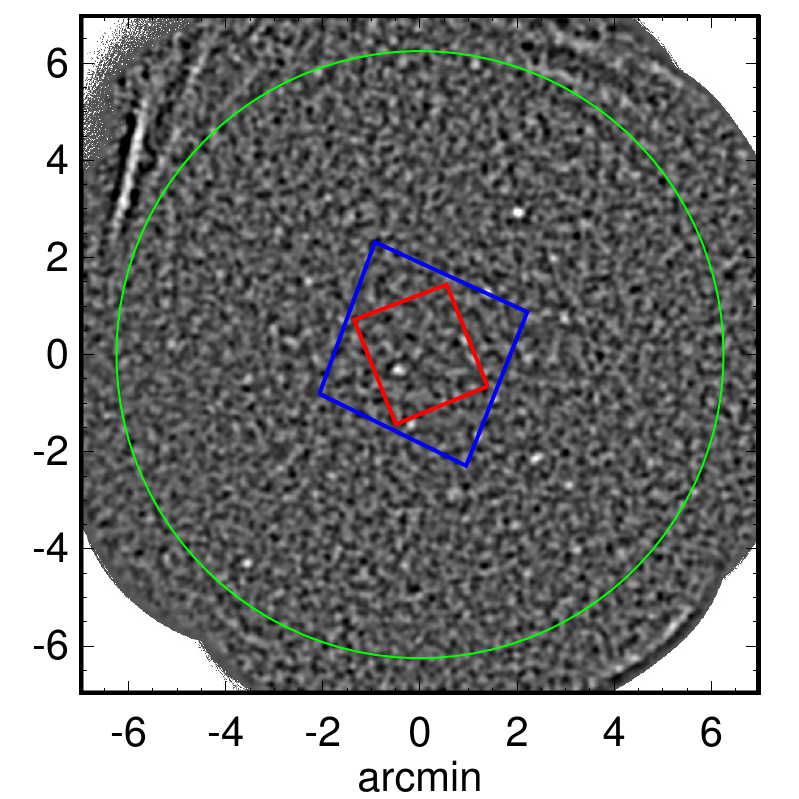}
\includegraphics[width=6cm]{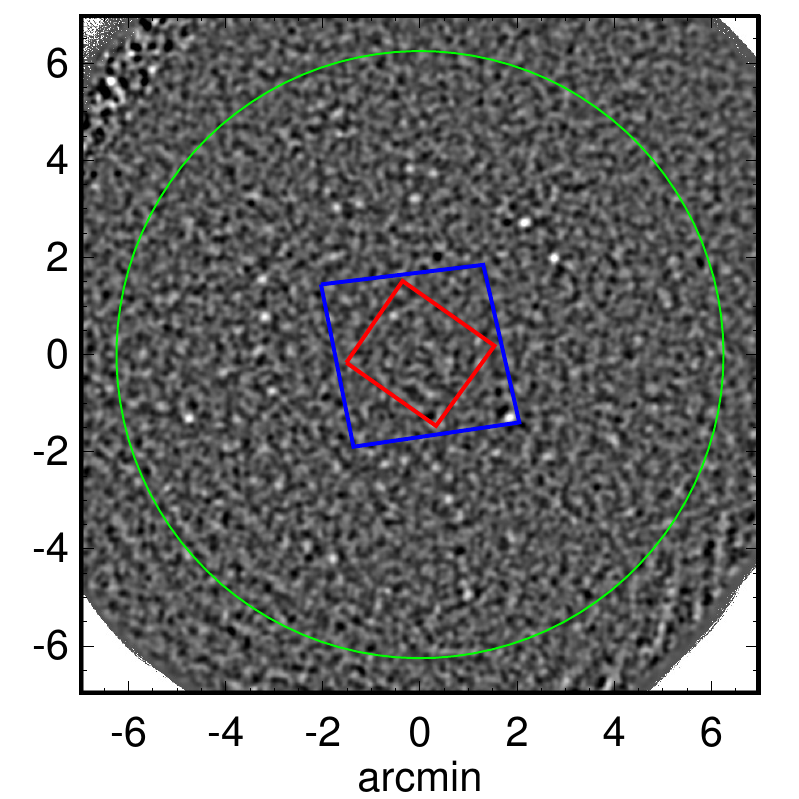} 
\includegraphics[width=6cm]{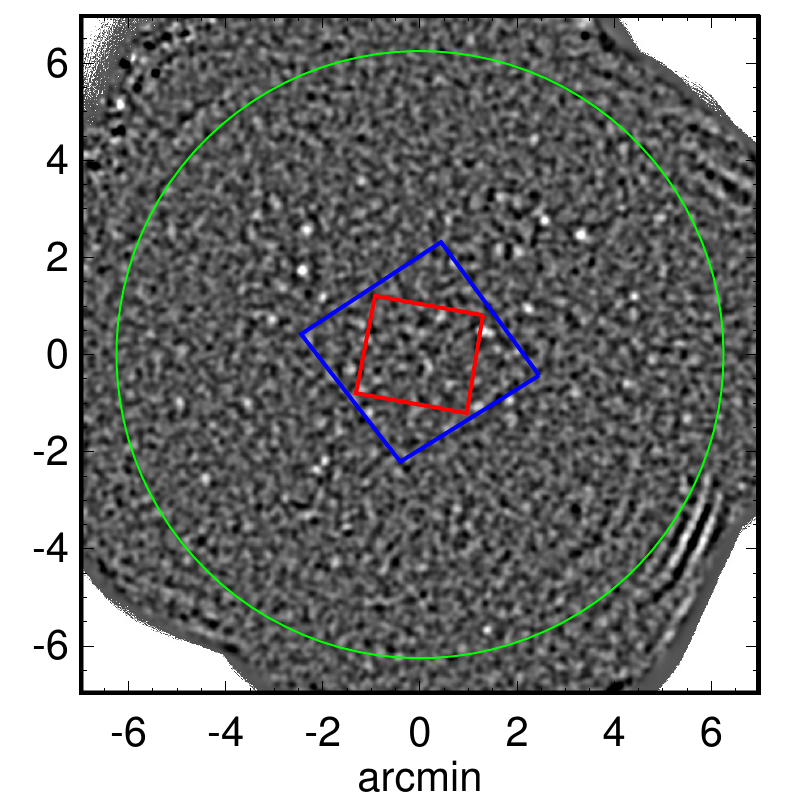}
\caption{850 $\mu$m (top) and 450 $\mu$m (bottom) S/N maps of the three Frontier Fields, MACS\,J0416.1--2403, MACS\,J0717.5+3745 and MACS\,J1149.5+2223. The blue and red boxes represent, respectively, the positions of ACS and WFC3 coverage for the Frontier Field program. The green circles show the coverage of our VLA 6 GHz observations. Our 3 GHz images cover four times the area of the 6 GHz observations. In this work, we focus on the areas of {\it HST} coverage.}
\label{fig:figure1}
\end{figure*}

Following \cite{Chen2013Faint-Submillim,Chen2013Resolving-the-C} and \cite{Hsu2016The-Hawaii-SCUB}, we reduced the data using the Dynamic Iterative Map Maker (DIMM) in the SMURF package from the STARLINK software \citep{2013MNRAS.430.2545C}. DIMM performs pre-processing and cleaning of the raw data (e.g., down-sampling, dark subtraction, concatenation, flat-fielding) as well as iterative estimations to remove different signals from astronomical signal and noise. We adopted the standard ``blank field'' configuration file, which is commonly used for extragalactic surveys to detect low signal-to-noise ratio (S/N) point sources. Please refer to \cite{Hsu2016The-Hawaii-SCUB} for a detailed description of our SCUBA-2 data reduction and calibration. In Figure~\ref{fig:figure1}, we show the 850 $\mu$m S/N maps for the three cluster fields with the regions of {\it HST} and VLA 6 GHz observations overlaid.

\begin{table*}
\caption{Summary of JCMT/SCUBA-2 observations}
\begin{center}
\begin{tabular}{cccccccc}
\hline \hline	

 Field  & R.A. & Decl. & Redshift  & Weather & Exposure   & ${\sigma}$\footnote{Central 1$\sigma$ sensitivity of the map at 450 and 850 $\mu$m. These are the statistical/instrumental noise values 
 directly from the reduced rms maps. Therefore, the effect of confusion noise is not included.}    \\
           &       &     &      &        &   (hr)    &   (mJy beam$^{\rm -1}$)    \\

\hline

MACS\,J0416.1--2403     & 04 16 08.9  &  $-$24 04 28.7     & 0.396  &1+2+3 &  24.0+1.0+4.0   &  [2.31,0.36]  \\  
MACS\,J0717.5+3745     &  07 17 34.0 &  \,\,\,\,\,37 44 49.0 & 0.545 &1+2+3 &  30.7+8.0+1.5 &  [2.03,0.34]  \\      
MACS\,J1149.5+2223     & 11 49 36.3  &  \,\,\,\,\,22 23 58.1 & 0.543  &1+2+3 &  29.0+3.5+3.4  &  [1.63,0.30]  \\ 

\hline \hline  
 
\end{tabular}  
\label{tab:table1}
\end{center}
\end{table*}

\subsection{VLA Images}

The VLA observations\footnote{Project codes: 14A-012, 15A-282} were carried out in the A (maximum baseline = 36.4 km) and the C (maximum baseline = 3.4 km) configurations using both the S-band (2-4 GHz) and C-band (4-8 GHz) receivers. For the S band, two 1 GHz Intermediate Frequency (IF) band pairs were used, both with right- and left-hand circular polarization, and sampled at 8 bits, while for the C band two 2 GHz IF band pairs were utilized with 3 bit sampling. The 1 and 2 GHz-wide bands were then divided by the WIDAR correlator into 8 and 16 128 MHz wide spectral windows, respectively, each with 64 spectral channels and four polarization products (RR, LL, RL, and LR). The on-source 
integration times for each of the three targets, in each of these band / configuration pairs are given in Table~\ref{tab:obs}, along with the equatorial coordinates of the targets themselves. The primary and secondary calibrators are also listed 
for each target.

\begin{table*}
\begin{center}
\begin{minipage}{140mm}
\caption{Summary of VLA observations, including calibrators and the on-source integration times for each configuration/band pairings. }
\begin{tabular}{ccccccc} 
\hline \hline
                           &           &  Band & Config   & MACS\,J0416.1--2403 & MACS\,J0717.5+3745 & MACS\,J1149.5+2223 \\ \hline
Primary cal         &           &           &              & 3C48 & 3C147 & 3C286  \\
Secondary cal    &           &  S       &              & J0416-1851 & J0714+3534 & J1158+2450 \\
Secondary cal    &           &  C       &              & J0416-1851 & J0714+3534 & J1150+2417  \\

Integration time & (hr)      & S & A    & 35.2       & 25.2           & 27.3      \\
Integration time & (hr)      & S & C   & 1.92       & 1.52           & 1.57       \\
Integration time & (hr)      & C & A    & 18.8       & 13.7           & 13.9      \\
Integration time & (hr)      & C & C   & 1.05      &  0.56            & 0.62      \\  
\hline \hline

\end{tabular}
\label{tab:obs}
\end{minipage}
\end{center}
\end{table*}

The data of each individual observing session were initially processed using the NRAO VLA pipeline\footnote{{\href{https://science.nrao.edu/facilities/vla/data-processing/pipeline}{https://science.nrao.edu/facilities/vla/data-processing/pipeline}}}. This is a set of scripts for the Common Astronomy Software Applications \citep[{\sc CASA}\footnote{\href{http://casa.nrao.edu/}{http://casa.nrao.edu/}};][]{McMullin2007CASA-Architectu} package designed to perform basic calibration steps on continuum data for total intensity (Stokes I) science. After Hanning-smoothing, the pipeline performs various data editing steps such as the flagging of data due to antenna shadowing, visibilities with amplitudes that are exactly zero, and integrations when the antennas are not 
one-source. A first pass of radio frequency interference (RFI) excision from the calibrator and target scans is performed using a sliding window statistical filter. The pipeline also performs delay and bandpass calibration using the primary calibrators.
Time-dependent antenna-based complex gain corrections are derived using the secondary calibrator and interpolated for application to the target scans. A gain correction is derived independently for each spectral window.

Following the execution of the pipeline, the target field from each pointing was split into a single measurement set. The {\sc CASA} {\sc mstransform} task was then used to add a {\sc WEIGHT\_SPECTRUM} column to the visibilities. This column has the same shape as the {\sc DATA} column and allows a unique weight to be assigned to each visibility point for use in subsequent imaging. The {\sc statwt} task was then used to adjust values in the {\sc WEIGHT\_SPECTRUM} based on the time-dependent statistical properties of the visibilities for each baseline. This step (often) proves to be effective at suppressing low-level RFI or other issues with the data that are missed by the automated flagging routines.

The target fields were then imaged using the {\sc wsclean} software \citep{Offringa2014} and Briggs weighting (robustness parameter $=$ 0.2), producing images of 16,384~$\times$~16,384 pixels, with pixel sizes of 0$\farcs$1 and 0$\farcs$06 for S and C bands, respectively. Images were produced for each band and each cluster by jointly gridding and deconvolving all of the relevant measurement sets. Spectral behavior of the sources (both intrinsic toward the beam center, and instrumentally perturbed off-axis) was captured during deconvolution by imaging the data in four spectral sub-bands across the band. The approach used by {\sc wsclean} during deconvolution is to find peaks in the full-band image and then deconvolve these in each sub-band independently. For major-cycle purposes, clean components were fitted by a second order polynomial when predicting the visibility model. Cleaning was terminated after 100,000 iterations or when a the peak pixel in the full-band residual map reached a threshold of 1.0~$\mu$Jy, whichever occurred sooner. Imaging concludes with the model being restored into the full-band residual map, using a 2D Gaussian as fitted to the main lobe of the point spread function as the restoring beam. The data from both configurations (A and C) were combined during deconvolution and imaging. In Table~\ref{tab:table3}, we provide the synthesized beams of the images for both bands. We caution that the small beams of these images might resolve out some extended emission and 
therefore miss some sources.

The primary beam sizes (HPBW) are $\sim$ 14$'$ at S band and $\sim$ 7$'$ at C band. Primary beam correction was applied to the final image by dividing it by a model of the VLA Stokes-I beam at the band center. The model itself was obtained by running the {\sc CASA clean} task and using the predicted sensitivity ({\sc .flux}) image. This is a somewhat crude approach for data with such a large fractional bandwidth; however (1) primary beam correction via projection-based gridding is not yet viable, and (2) the band center beam model differs from the zeroth-order Taylor-term beam model predicted by the {\sc widebandpbcor} task by a couple of percent at most, so for our purposes the approaches are essentially equivalent.

The data reduction at this stage is designed to provide an initial set of Stokes-I images at S and C bands. Improvements in the imaging is possible via self-calibration techniques and this work is on-going. In-band and dual band (S--C) spectral index maps will be produced once the calibration is finalized. The observations were also scheduled to allow polarimetric calibration, and this is also forthcoming.

\begin{table*}
\caption{Synthesized beams of the 3 and 6 GHz images for each field.}
\begin{center}
\begin{tabular}{ccccccc}
\hline \hline	

           &                &   {3 GHz} &     &     & {6 GHz}   &       \\
 Field  & $b_{\rm max}$ & $b_{\rm min}$ &  $b_{\rm PA}$  & $b_{\rm max}$ & $b_{\rm min}$ &  $b_{\rm PA}$ \\

\hline

MACS\,J0416.1--2403  &   0$\farcs$81 & 0$\farcs$42 &  0.65$^{\circ}$ &  0$\farcs$47 & 0$\farcs$24  &  20.96$^{\circ}$   \\  
MACS\,J0717.5+3745   &  0$\farcs$49 & 0$\farcs$44 &  78.29$^{\circ}$ &  0$\farcs$27 & 0$\farcs$23  &  -76.10$^{\circ}$ \\     
MACS\,J1149.5+2223   &  0$\farcs$44 & 0$\farcs$41 &  33.38$^{\circ}$ &  0$\farcs$24 & 0$\farcs$22  &  46.37$^{\circ}$ \\   

\hline \hline 
 
\end{tabular}  
\label{tab:table3}
\tablecomments{$b_{\rm max}$, $b_{\rm min}$, and $b_{\rm PA}$ represent the major axis FWHM (in arcsecs), minor axis FWHM (in arcsecs), and position angle (in degrees), respectively.}
\end{center}
\end{table*}

\subsection{HST Images and Photometry}

We retrieved the {\it HST} Frontier Fields images and the {\it HST} images from the Cluster Lensing And Supernova survey with Hubble (CLASH; \citealt{Postman2012The-Cluster-Len}) 
archive\footnote{https://archive.stsci.edu/prepds/clash/} for the passbands that are not included in the Frontier Fields program. We ran SExtractor \citep{Bertin1996SExtractor:-Sof} in dual-image mode using F814W and F160W as detection bands to produce two sets of photometric catalogs. The deblending parameters {\sc deblend\_nthresh} and {\sc deblend\_mincont} were set to be 32 and 0.005, respectively. For a source that is within the WFC3 coverage, we use the F160W-detected photometry instead of the F814W-detected one.

\subsection{Spitzer Images and Photometry}\label{sec:spitzer}

We retrieved the {\it Spitzer} Frontier Fields data at 3.6 and 4.5 $\mu$m, and we used the {\it Spitzer} image processing package {\sc mopex} \citep{Makovoz2005Mosaicking-with,Makovoz2005Point-Source-Ex,Makovoz2006MOPEX:-a-softwa} to extract sources. Photometry estimation and deblending were performed by the default Point Response Function (PRF) fitting algorithm.

\subsection{Other Ancillary Data and Photometric Catalogs}

\cite{Brammer2016Ultra-deep-K-S-} recently provided deep, calibrated $K_s$-band images of all six of the Hubble Frontier Fields using the instruments HAWK-I on the VLT and 
MOSFIRE on the Keck I telescope. We retrieved the images of our three cluster fields and ran SExtractor to perform source extraction. We also obtained the images and photometric catalogs from the 
CLASH archive that were obtained with Supreme-Cam on the Subaru telescope, WIRCam and MegaCam on the Canada--France--Hawaii telescope (CFHT), as well as the source 
catalogs of {\it Herschel} PACS and SPIRE passbands from the {\it Herschel Lensing Survey} \citep{Rawle2016A-complete-cens}.

\section{Sample Selection}

\subsection{SCUBA-2 Source Extraction}\label{sec:extraction}

In this work, we focus on 850 $\mu$m selected sources, and we detected sources down to a 4$\,\sigma$ level. \cite{Casey2013Characterisatio} and \cite{Chen2013Faint-Submillim,Chen2013Resolving-the-C} have shown that sources detected above a 4$\,\sigma$ level from their SCUBA-2 maps (with central 1$\,\sigma$ sensitivity of $\sim$ 0.8 mJy at 850~$\mu$m) have a contamination rate of $\leq$ 5\%. Following \cite{Hsu2016The-Hawaii-SCUB}, we estimate the contamination rate by constructing the source-free "jackknife maps" at both wavelengths. A jackknife map is a pure noise image with sources removed; it is created by subtracting one-half of the data from the other, then scaling each pixel value by a factor of $\sqrt{t_1 \times t_2}/(t_{1}+t_{2})$, with $t_1$ and $t_2$ representing the integration time of each pixel from the two halves of data. In each field, we searched for 4$\,\sigma$ sources within the area where the noise values are less than three times the central noise (an area of $\sim$ 130 arcmin$^2$). The ratio of the total number of sources from the jackknife maps and from the science maps is 10/292, or 3.4\%. If we only consider the regions that are covered by {\it HST}/ACS, the ratio is 1/44, or 2.3\%.

To perform source extraction, we generated the PSFs by averaging all the primary calibrators. Following the methodology of source extraction in \cite{Chen2013Faint-Submillim,Chen2013Resolving-the-C} and \cite{Hsu2016The-Hawaii-SCUB}, we identified the pixel with the maximum S/N, subtracted this pixel and its surroundings using the PSF centered and scaled at the position and value of this pixel, and then searched for the next maximum S/N. We iterated this process until the detection threshold was hit. We only selected sources in the areas covered by the {\it HST} Frontier Fields since these sources are the most highly lensed ones with deep {\it HST} photometry. There are 44 850 $\mu$m sources within the {\it HST} coverage. We then measured the 450 $\mu$m flux density of each 850 $\mu$m source by searching for the 
maximum 450 $\mu$m peak flux within the 850 $\mu$m beam.

\subsection{Flux Deboosting}

The flux densities we measured from the SCUBA-2 maps are boosted by both Eddington bias \citep{Eddington1913On-a-formula-fo} and confusion noise \citep{Condon1974Confusion-and-F}. 
In addition, the flux errors we obtained are purely from statistical/instrumental noise, which does not include the above effects. We therefore need to run Monte Carlo simulations to
correct the measured flux densities and their uncertainties. To perform such simulations, one normally uses a number counts model to populate sources onto the jackknife 
map, runs source extraction on this map, and then compares the input and measured flux densities. In a lensed field, however, it is tricky to run these simulations because 
assumptions of a lens model and a source plane redshift are required to project the simulated sources onto the image plane and to magnify their fluxes. 

Here we used a simpler but rather time-consuming method that does not require lens modeling. We added three sources into our science map with random flux densities at random 
positions at a time and then performed source extraction, recording the input and measured flux densities of the sources we inserted. We only used the central area of each map that is 
covered by the {\it HST}. Only a small number of sources were inserted at a time in order to avoid over-crowdedness, such that the flux measurements for these sources are not influenced 
by each other. We repeated this procedure until we obtained 50000 pairs of input and measured flux densities. An inserted source is considered to be recovered if it is detected within 
the HWHM of the SCUBA-2 beam from the original input position. 
    
 In Figure~2, we show the boosting factor as a function of detection S/N at 450 and 850 $\mu$m for MACSJ0416. The boosting factor is measured as the ratio of the measured and input 
 flux densities. The red line in each panel of Figure 2 represents the median boosting factor, and the two blue lines enclose the 1$\,\sigma$ spread. We deboosted the flux densities of our sources 
 in each map using the median boosting factor and the corresponding 1$\,\sigma$ uncertainty.

\subsection{Confusion Limit at 850 $\mu$m}

To estimate the confusion limit of SCUBA-2 observations at 850 $\mu$m, we use the formalism of \cite{Condon1974Confusion-and-F} and define the beam size as $\Omega_{\rm b} = \pi({\rm FWHM}/2.35)^2$, where FWHM is 14$\farcs$5. Following \cite{Hogg2001Confusion-Error}, an image is considered confused when the source density exceeds one source per 30 beams. We adopt a broken power law for the differential 
number counts 
\begin{equation}                                                               
\frac{dN}{dS}=\left\{\begin{matrix}
N_0\left ( \frac{S}{S_0} \right )^{-\alpha} ~~{\rm if}~S\leq S_0 \\ 
N_0\left ( \frac{S}{S_0} \right )^{-\beta} ~~{\rm if}~S > S_0
\end{matrix}\right.
\end{equation}  
We can then calculate the cumulative counts, $N(>S)$, which represents the number density of sources that are brighter than $S$. Equating $N(> S)$ and $1/30 \Omega_{\rm b}$ leads 
to the confusion limit of blank fields
\begin{equation}
\label{eq:conf}
S_{\rm c} = \left \{ \frac{\alpha -1}{N_{0} S_{0}^{\alpha}}  \left [ \frac{1}{30 \Omega_{\rm b}}  + N_0 S_0 \left (\frac{1}{\alpha -1} - \frac{1}{\beta -1} \right ) \right ] \right \}^\frac{1}{1 - \alpha}
\end{equation}
Adopting the best-fit parameters of the broken power law in \cite{Hsu2016The-Hawaii-SCUB}, we obtain $S_{\rm c} = 1.64$ mJy.

In a lensed field, the cumulative number counts become $N_{\rm lens}(> S) = N(>S/ \mu) / \mu$, where $\mu$ is the lensing magnification and $S$ is the observed flux density. Equating $N_{\rm lens}(> S)$ and $1/30 \Omega_{\rm b}$, we obtain the observed confusion limit, $S_{\rm c}$ (on the image plane), as a function of $\mu$. The confusion limit on the source plane is $S_{\rm c}/ \mu$. The observed 
confusion limit is higher than that of blank fields, but on the source plane it is lower. For example, $S_{\rm c}$ and $S_{\rm c}/\mu$ are $\sim$ 1.93 
and 0.96, respectively, for $\mu = 2$.

Since we have detected sources down to a deboosted flux density of $\sim $ 1.4 mJy, many of them should be close 
to or below the observed confusion limit (which depends on $\mu$ and therefore position). As a consequence, we caution that our source 
detection at $S_{\rm 850 \mu m} \lesssim$ 2 mJy is not complete because there must be sources that we missed due to source confusion.

\subsection{VLA Source Extraction}\label{sec:vlaextraction}

Given the higher detection rate at 3 GHz than at 6 GHz, we used the 3 GHz images to search for the counterparts to our 850~$\mu$m sources. To extract the flux densities and positions of the 3 GHz sources, we first identified all the pixels that are local maxima and have S/N $\geq$ 5. We took the values of these pixels as the peak fluxes (per beam) of the sources. We then used the {\sc CASA} {\sc imfit} task to fit 2D Gaussian functions to measure another set of flux densities. Note that some of the flux densities measured with this method have S/N $<$ 5, but we still keep these sources given their well detected peak fluxes.

Using the source positions at 3 GHz as prior, we searched for the 6 GHz counterparts with 5$\,\sigma$-detected peak fluxes. For the sources that are not detected at 6 GHz, we measured their 5$\,\sigma$ limits 
at the 3 GHz positions. We again used {\sc imfit} to measure a second set of flux densities for the detected 6 GHz sources. In addition, we measured another set of flux densities with {\sc imfit} from the 6 GHz 
images that were convolved to match the beams of our 3 GHz images, using the {\sc CASA} {\sc imsmooth} task. This set of measurements along with the 3 GHz flux densities will be used to compute the spectral 
indices of our radio sources (Section~\ref{sec:radio}). We adopted this procedure to counter the beam/resolution difference between the two bands.

Further discussion about the source extraction as well as its completeness and contamination rate will be presented in the upcoming paper, Heywood et al. (2017, in preparation).

\begin{figure*}
\begin{center}    
\includegraphics[width=7cm]{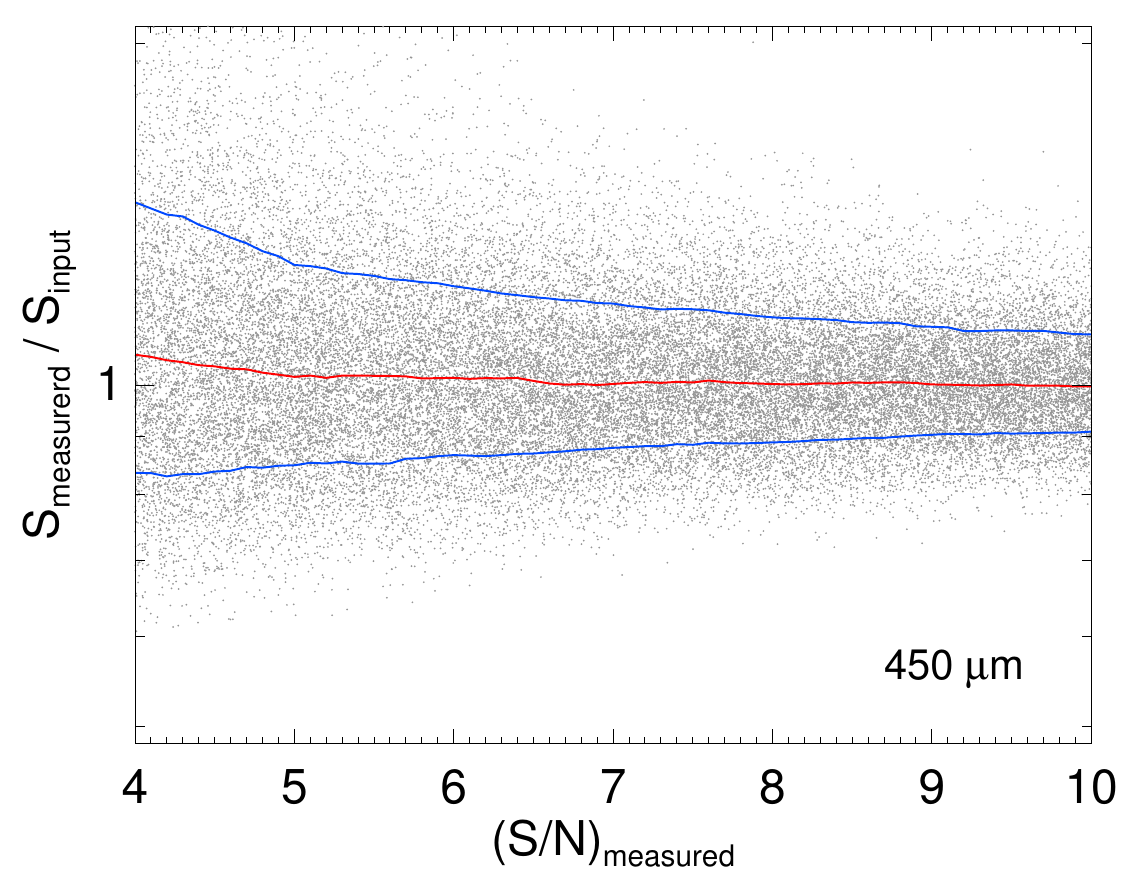}
\includegraphics[width=7cm]{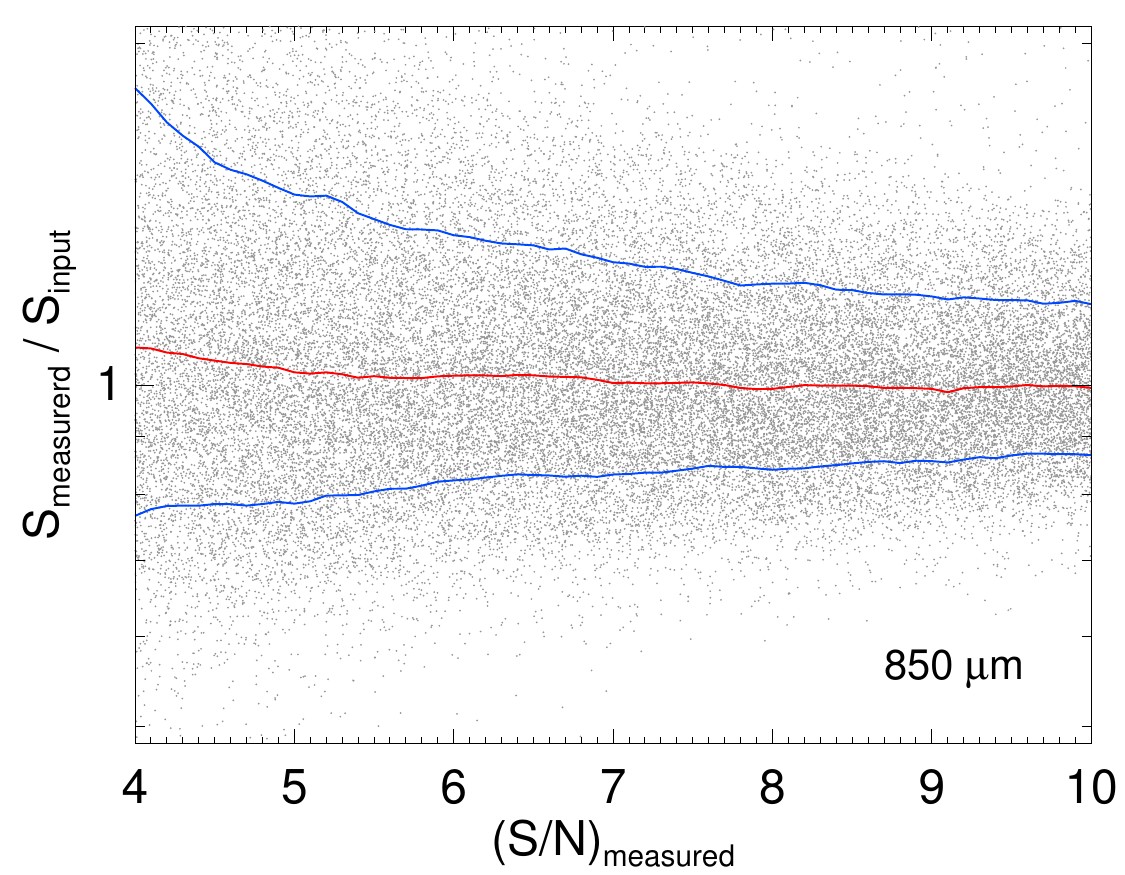}
\caption{Boosting factor as a function of detection S/N at 450 $\mu$m (left) and 850 $\mu$m (right) for MACS\,J0416.1--2403 from our Monte Carlo simulations. The boosting factor is 
measured as the ratio of the measured and input flux densities. The red line in each panel represents the median boosting factor, and the two blue lines enclose the 1$\,\sigma$ spread. We deboosted 
the flux densities of our sources in each map using the median boosting factor and the corresponding 1$\,\sigma$ uncertainty.}
\end{center}
\label{fig:figure2}
\end{figure*}

\subsection{Counterpart Identification at 3 GHz}\label{sec:counterpart}

We used the corrected-Poissonian probability \citep{Downes1986The-Parkes-sele}, the so-called $p$-values, as well as redshift cuts to perform counterpart identification. A radio 
source is considered the counterpart to the SCUBA-2 source if $p < 0.05$ and its redshift is higher than the corresponding cluster redshift. The $p$-value is defined 
as $p = 1 - exp(\pi n \theta^2)$, where $n$ is the radio source density\footnote{$n = 6.1 \times 10^{-4}$~arcsec$^{-2}$, $6.0 \times 10^{-4}$~arcsec$^{-2}$, and $8.8 \times 10^{-4}$~arcsec$^{-2}$ 
for MACSJ0416, MACSJ0717, and MACSJ1149, respectively.} and $\theta$ is the offset between the radio and the submillimeter sources. We adopted the redshift cuts as well because 
galaxies at redshifts lower than the cluster redshifts are not the lensed, faint galaxies that we are interested in. Besides, 850 $\mu$m selected SMGs have been mostly found at 
$z > 0.5$ \citep{Chapman2003A-median-redshi,Chapman2005A-Redshift-Surv}. Therefore, those low-redshift galaxies are most likely just random radio sources that are not 
associated with our SCUBA-2 sources.

A total of 17 radio sources have $p < 0.05$, four of which have spectroscopic redshifts (\citealt{Ebeling2014Spectroscopic-R,Schmidt2014Through-the-Loo,Grillo2015CLASH-VLT:-Insi,Treu2015The-Grism-Lens-,Balestra2016CLASH-VLT:-Diss}). We used the BPZ code (Bayesian photometric redshift estimation; \citealt{Benitez2000Bayesian-Photom}) and the default galaxy templates \citep{Benitez2004Faint-Galaxies-,Coe2006Galaxies-in-the} to compute the photometric redshifts of those galaxies without spectroscopic redshifts. We fitted the templates to the {\it HST} photometry using isophotal magnitudes 
and obtained robust photometric redshifts for the sources that are detected in at least four bands (but not four ACS bands exclusively). One source in MACSJ0717 (0717-1 in Table~\ref{tab:table4}) is only covered in three {\it HST}/ACS bands but is covered in the Subaru and CFHT images from CLASH. For this source, we simply took the photometric redshift ($z = 1.14^{+0.07}_{-0.10}$) from the CLASH Subaru catalog\footnote{Note that, in \cite{Rawle2016A-complete-cens}, the photometric redshift ($z=0.89$) from the CLASH {\it HST} catalog is used for this source. We chose the value from the CLASH Subaru catalog because this source is detected in only two {\it HST} bands of CLASH but detected in six Subaru/CFHT bands.}. We corrected all of the magnitudes for Galactic dust extinction from \cite{Schlafly2011Measuring-Redde} before running BPZ.

For the sources without spectroscopic and photometric redshifts (0416-4 and 0717-2 in Table~\ref{tab:table4}), we used their 450 $\mu$m-to-850 $\mu$m flux ratios to crudely 
estimate the source redshifts. We converted the flux ratios to redshifts using a modified blackbody spectral energy distribution (SED) of the form $S_{\nu} \propto (1 - e^{-\tau(\nu)})B_{\nu}(T)$, where 
$\tau(\nu) = (\nu / \nu_0)^{\beta}$ and $\nu_0 =3000$ GHz, assuming $\beta = 1.5$ and dust temperature of 41.2 K. We chose this value of the dust temperature based on the results 
of our model fits to the FIR photometry, which we will describe in Section~\ref{sec:FIR}.

We removed two of these 17 radio sources with $p < 0.05$ because of their low redshifts. Our final sample therefore consists of 14 SCUBA-2 sources and 15 3 GHz 
counterparts (one doublet). Only five of the SCUBA-2 sources (six of the radio counterparts) are detected at 6 GHz with peak fluxes above a 5$\,\sigma$ level. We show the 3 GHz 
and {\it HST}/ACS images of this final sample in Figure~3. Table~\ref{tab:table4} gives the coordinates, redshifts, observed SCUBA-2 and radio flux densities of these galaxies. It is 
possible that a radio counterpart is not responsible for the total emission of the corresponding 850~$\mu$m source, given that blended multiples are common for single-dish submillimeter sources (e.g., \citealt{Wang2011SMA-Observation,Barger2012Precise-Identif,Smolcic2012Millimeter-imag,Hodge2013An-ALMA-Survey-,Bussmann2015HerMES:-ALMA-Im,Simpson2015The-SCUBA-2-Cos}). However, in this work, 
we assume that these 15 radio sources contribute the total 850~$\mu$m emission, in order to 
derive the IR properties of our 14 SCUBA-2 sources (Section~\ref{sec:FIR}). Submillimeter/millimeter interferometry is the only method to find out whether there are other radio-faint counterparts to these SCUBA-2 sources.

\cite{Cowie2017A-Submillimeter} inspected the positional uncertainty of the SCUBA-2 850~$\mu$m sources detected in the {\it Chandra} Deep Field-North (CDF-N). They found that the offsets 
between the SCUBA-2 positions and the SMA positions or the 1.4 GHz counterparts are all $<$ 4$\farcs$5 for 102 sources with 850 $\mu$m flux densities above 2 mJy. In our final sample, only 0717-4 has 
an offset of $\sim$ 5$\farcs$1 between the radio and submillimeter positions, and the others have offsets of $<$ 4$\farcs$5. Besides, $p=0.05$ essentially corresponds to matching radii of $\sim$ 5$\farcs$2, 
5$\farcs$2, and 4$\farcs$3 for MACSJ0416, MACSJ0717, and MACSJ1149, respectively. Our results are therefore quite consistent with what Cowie et al. found. If we chose a fixed 
matching radius of 4$\farcs$5 and the same redshift cuts for counterpart matching, we would obtain a sample of 14 SCUBA-2 sources with only one source different from 0717-4.

\subsection{Lens Models}\label{sec:lens}

In order to compute the intrinsic flux densities, luminosities, and SFRs of our lensed SMGs, the lensing magnifications are required, which depend on both the source redshifts and the lens models 
of the clusters. A set of lens models from eight independent teams are available for the {\it HST} Frontier Fields. These teams include Bradac \citep{Bradac2005Strong-and-weak,Bradac2009Focusing-Cosmic,Hoag2016The-Grism-Lens-}, CATS \citep{Jullo2009Multiscale-clus,Jauzac2012A-weak-lensing-,Jauzac2014Hubble-Frontier,Jauzac2015MNRAS.452.1437,Jauzac2015Hubble-Frontier,2014MNRAS.444..268R}, Diego \citep{Diego2005Non-parametric-,Diego2005-A1689,Diego2007Combined-recons,Diego2015}, GLAFIC \citep{Oguri2010The-Mass-Distri,Kawamata2016Precise-Strong-}, Merten \citep{Merten2009Combining-weak-,Merten2011Creation-of-cos}, Sharon \citep{Jullo2007A-Bayesian-appr,Johnson2014Lens-Models-and}, Williams \citep{Liesenborgs2006A-genetic-algor,Mohammed2014,Grillo2015CLASH-VLT:-Insi,Sebesta2016}, 
and Zitrin \citep{Zitrin2009New-multiply-le,Zitrin2013CLASH:-The-Enha}. 

Following \cite{Coe2015Frontier-Fields}, we estimated the median and 68.3\% range of the magnification values from the full range of each model in the {\it HST} Frontier Fields 
archive\footnote{https://archive.stsci.edu/prepds/frontier/lensmodels/} as well as the uncertainties of the source redshifts. The Merten models are excluded in our estimations because 
they have much lower resolution (8$\farcs$33). For each field, we used the newest model from each team. However, because different models have different spatial coverage, different sources in the 
same field are often not covered by the same amount of lens models. 

In order to be consistent for the sources in the same field, for MACSJ0416 and MACSJ0717, we only used the models that cover all the sources in each field. However, for MACSJ1149, only the CATS and Sharon models 
can cover all the six sources, which would result in much smaller systematic errors of magnifications than those of the sources in the other two fields. Ideally, at least three models should 
be included to examine the systematic effect on the magnifications. We therefore used a set of four models, which all cover 1149-1, 1149-2, 1149-3, 1149-4, and 1149-6, and a different set 
of three models (all the available models) for 1149-5. In Table~\ref{tab:table5}, we tabulate the models we included and the resulting lensing magnification for each of our sources.

\begin{table*}
\caption{Coordinates and redshifts, as well as SCUBA-2 and VLA flux densities of the 3 GHz identified sample of 850 $\mu$\MakeLowercase{m} sources}
\begin{center}
\begin{tabular}{ccccccccccc}
\hline \hline	
 ID  & R.A. & Decl. &  Redshift  & $S_{850 \mu{\rm m}}$ 
 & $S_{450 \mu{\rm m}}$ & $S_{\rm 3GHz,peak}$ & $S_{\rm{3GHz,fit}}$ & $S_{\rm 6GHz,peak}$ & $S_{\rm{6GHz,fit}}$ &  $S_{\rm{6GHz-cv,fit}}$\\ 
           &       &        &      &  (mJy)   & (mJy)  &  ($\mu$Jy beam$^{-1}$) & ($\mu$Jy) & ($\mu$Jy beam$^{-1}$) & ($\mu$Jy) & ($\mu$Jy) \\
\hline
0416--1    & 04 16 10.80 & $-$24 04 47.6 &   2.087 (s)  & 4.38 $\pm$ 0.72  &  14.5 $\pm$ 3.5  & 8.0 $\pm$ 1.0    & 14.7 $\pm$ 3.3 &$< 4.3$  &  ... & ... \\ 
0416--2    & 04 16 13.23 & $-$24 03 19.8 &   0.9063 (s) & 2.22 $\pm$ 0.79  & 7.5 $\pm$ 3.5    & 12.2 $\pm$ 1.1  & 27.8 $\pm$ 4.1 &$< 5.6$  &  ... & ... \\ 
0416--3    &      ...               &      ...            &      ...           &  2.18 $\pm$ 0.79 & 14.1 $\pm$ 4.0 &           ...               &            ...              &     ...       &  ... & ... \\      
0416--3--1   & 04 16 09.68 & $-$24 05 55.4 &  0.99 $\pm$ 0.10 (p)  &    ...    &   ...               & 31.1 $\pm$ 1.1  & 55.4 $\pm$ 3.7 &10.4 $\pm$ 1.1 & 20.8 $\pm$ 3.5 & 32.9 $\pm$ 4.6\\    
0416--3--2   & 04 16 09.64 & $-$24 05 55.2 &  1.01 $\pm$ 0.10 (p)  &    ...    &   ...               & 29.5 $\pm$ 1.1  & 31.9 $\pm$ 2.4 &13.0 $\pm$ 1.1 & 17.0 $\pm$ 2.4 & 18.5 $\pm$ 2.6\\        
0416--4    & 04 16 12.96  & $-$24 05 43.0 &  2.7$^{+2.7}_{-2.2}$ (r) & 1.78 $\pm$ 0.83  &  5.5 $\pm$ 3.4     & 11.8 $\pm$ 1.0 & 13.6 $\pm$ 2.1 &5.7 $\pm$ 1.0    & 9.0 $\pm$ 2.3  & 8.6 $\pm$ 2.3\\

0717--1    & 07 17 24.55  & \,\,\,\,\,37 43 29.7  &   1.14$^{+0.07}_{-0.10}$ (p)& 5.65 $\pm$ 0.88  & 18.7 $\pm$ 3.9 & 6.8 $\pm$ 0.9   & 18.2 $\pm$ 3.4  &$< 6.2$ &  ... & ... \\ 
0717--2    & 07 17 38.15  & \,\,\,\,\,37 46 17.0  &  4.5$^{+2.3}_{-1.3}$ (r)       & 3.25 $\pm$ 0.68  &  4.9 $\pm$ 2.9   & 5.6 $\pm$ 1.0   & 5.1 $\pm$ 1.5 &$< 5.5$    &  ... & ... \\      
0717--3    & 07 17 33.20  & \,\,\,\,\,37 44  01.5  &  1.54 $\pm$ 0.13 (p)  & 2.21 $\pm$ 0.61  &   7.2 $\pm$ 2.9          & 6.5 $\pm$ 1.1  & 15.2 $\pm$ 3.5  &$< 5.3$  &  ... & ... \\ 
0717--4    & 07 17 32.40  & \,\,\,\,\,37 43  19.7  &  0.78 $\pm$ 0.09 (p)  & 2.24 $\pm$ 0.69  &   6.9 $\pm$ 3.1          & 42.1 $\pm$ 1.0  & 48.9 $\pm$ 1.9 & 18.2 $\pm$ 1.0  &  24.3 $\pm$ 2.3  & 23.4 $\pm$ 1.6 \\ 

1149--1     & 11 49 30.66  & \,\,\,\,\,22 24 27.8  &   1.36 $\pm$ 0.12 (p)  & 4.75 $\pm$ 0.76  &  15.4 $\pm$ 2.7 & 5.2 $\pm$ 1.0   & 27.4 $\pm$ 8.3 &$< 4.9$  &  ... & ... \\ 
1149--2    & 11 49 36.07  & \,\,\,\,\,22 24 24.5  &  1.28 $\pm$ 0.11 (p)    & 2.20 $\pm$ 0.61  &   5.7 $\pm$ 2.3  & 6.6 $\pm$ 1.0   & 14.1 $\pm$ 4.0 &$< 5.1$  &  ... & ... \\       
1149--3    & 11 49 34.41  & \,\,\,\,\,22 24 45.3  &  0.9754 (s)                   & 2.23 $\pm$ 0.64  &  7.9 $\pm$ 2.6   & 14.4 $\pm$ 1.0 & 18.0 $\pm$ 2.0 & 8.0 $\pm$ 1.0 & 12.0 $\pm$ 2.3 & 12.4 $\pm$ 1.8\\
1149--4    & 11 49 35.47 &  \,\,\,\,\,22 22 32.0  &   1.24 $\pm$ 0.11 (p)    & 2.37 $\pm$ 0.69  &  4.1 $\pm$ 2.3  & 6.9 $\pm$ 1.0   & 12.5 $\pm$ 3.6 &$< 5.1$  &  ... & ... \\ 
1149--5    & 11 49 42.37 & \,\,\,\,\,22 23 39.6 & 1.56 $\pm$ 0.13 (p)       & 1.93 $\pm$ 0.74   &  8.1 $\pm$ 2.9  & 10.0 $\pm$ 1.0   & 12.1 $\pm$ 2.1 & 5.3 $\pm$ 1.0 & 9.1 $\pm$ 2.6 & 7.3 $\pm$ 1.5 \\    
1149--6   & 11 49 40.14 & \,\,\,\,\,22 22 33.4  &  0.93 $\pm$ 0.10 (p)      & 1.55 $\pm$ 0.77   &  3.1 $\pm$ 2.1  & 10.5 $\pm$ 1.0 & 17.2 $\pm$ 2.7 & $< 4.7$ &  ... & ... \\ 

\hline\hline 
\end{tabular}  
\label{tab:table4}
\tablecomments{R.A. and Decl. are the positions of 3 GHz flux peaks. Column 4: (s) stands for spectroscopic redshifts; (p) stands for photometric redshifts; and (r) stands for redshift estimates 
based on 450 $\mu$m-to-850 $\mu$m flux ratios. Columns 5 \& 6: deboosted SCUBA-2 flux densities. Columns 7 \& 9: peak fluxes; 5$\,\sigma$ limits are provided for the 
sources that are not detected at 6 GHz. Columns 8 \& 10: flux densities measured with a 2D Gaussian. Column 11: flux densities measured with a 2D Gaussian from the 6 GHz images 
convolved to match the beams of the 3 GHz images.}

\end{center}
\end{table*}

\begin{table*}
\caption{lens models used for each source and the resulting magnification value.} 
\begin{center}
\begin{tabular}{ccc}
\hline \hline	
 ID  & Magnification & Models \\ 
\hline 

0416--1   &  1.83$^{+0.54}_{-0.07}$  &  Bradac-v3~~CATS-v3.1~~Sharon-v3\\  
0416--2    & 1.67$^{+0.11}_{-0.14}$  &  '' \\               
0416--3    & 1.57$^{+0.26}_{-0.42}$  &  '' \\             
0416--4    & 1.21$^{+0.53}_{-0.08}$  &  '' \\    

0717--1    &  1.28 $\pm$ 0.16 &   Bradac-v1~~CATS-v1~~GLAFIC-v3~~Sharon-v2~~Zitrin-LTM-v1~~Zitrin-LTM-Gauss-v1\\   
0717--2    &  2.12$^{+0.42}_{-0.36}$ &    '' \\           
0717--3    &  3.04$^{+0.95}_{-0.43}$ &   '' \\         
0717--4    &  1.20$^{+0.09}_{-0.07}$ &   '' \\         

1149--1     &  1.64$^{+0.21}_{-0.35}$ & CATS-v1~~Sharon-v2.1~~Zitrin-LTM-v1~~Zitrin-LTM-Gauss-v1\\ 
1149--2    &  2.89$^{+0.66}_{-0.54}$  &  '' \\          
1149--3    &  3.08$^{+1.20}_{-0.44}$  &  '' \\    
1149--4    &  1.19$^{+0.20}_{-0.08}$  &   '' \\    
1149--5    & 1.44$^{+0.07}_{-0.18}$  &  CATS-v1~~Sharon-v2.1~~GLAFIC-v3~~~~~~~~~~~~~~~~~~~~~~~~~~~~~~~~~~~~~~~~~~\\        
1149--6    & 1.18$^{+0.10}_{-0.15}$  &  CATS-v1~~Sharon-v2.1~~Zitrin-LTM-v1~~Zitrin-LTM-Gauss-v1\\    
\hline \hline
\end{tabular}  
\label{tab:table5}
\tablecomments{The uncertainties of magnifications are propagated from the uncertainties of redshifts and lens models themselves. A '' sign means that the models are the same as the above. Note that for MACSJ0416, the 
CATS team provided two newest models (v3 and v3.1), which are based on the same data and method but a different amount of multiple images. Here we use the v3.1 model.}  
\end{center}
\end{table*}

\begin{figure*}

\includegraphics[width=2.95cm]{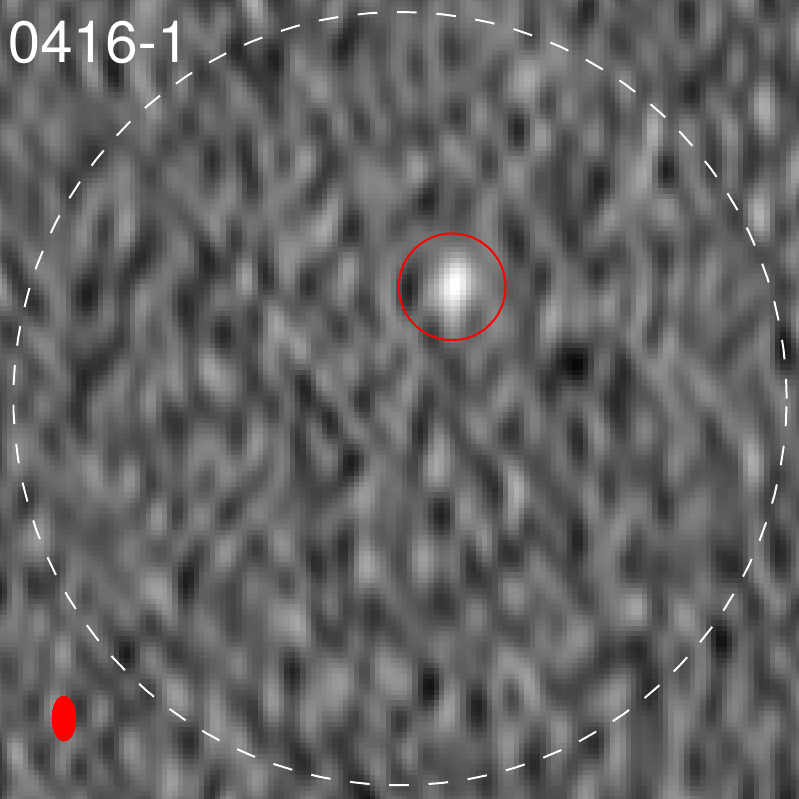}
\includegraphics[width=2.95cm]{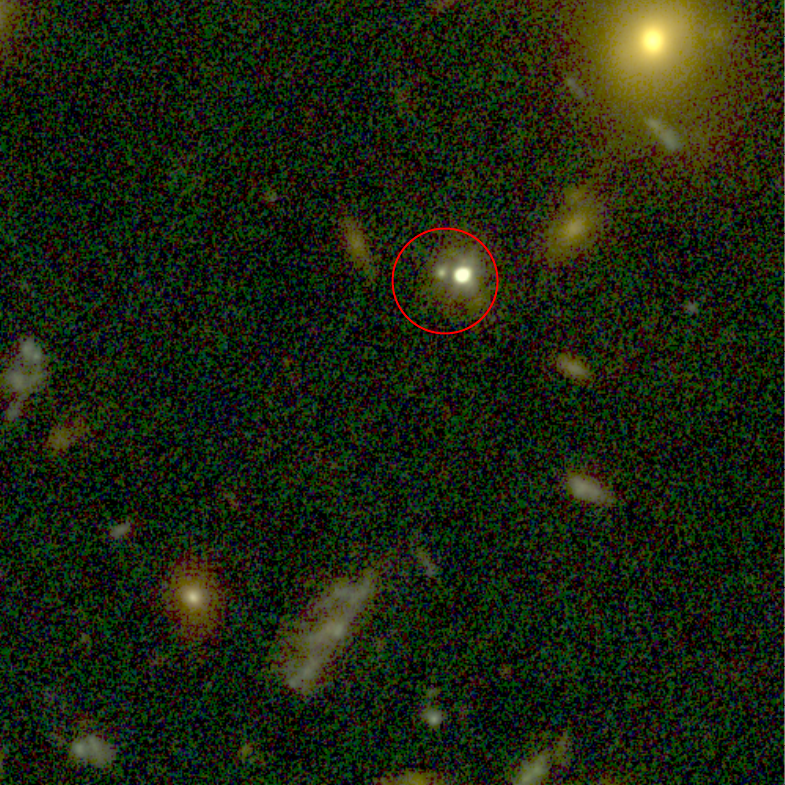} 
\includegraphics[width=2.95cm]{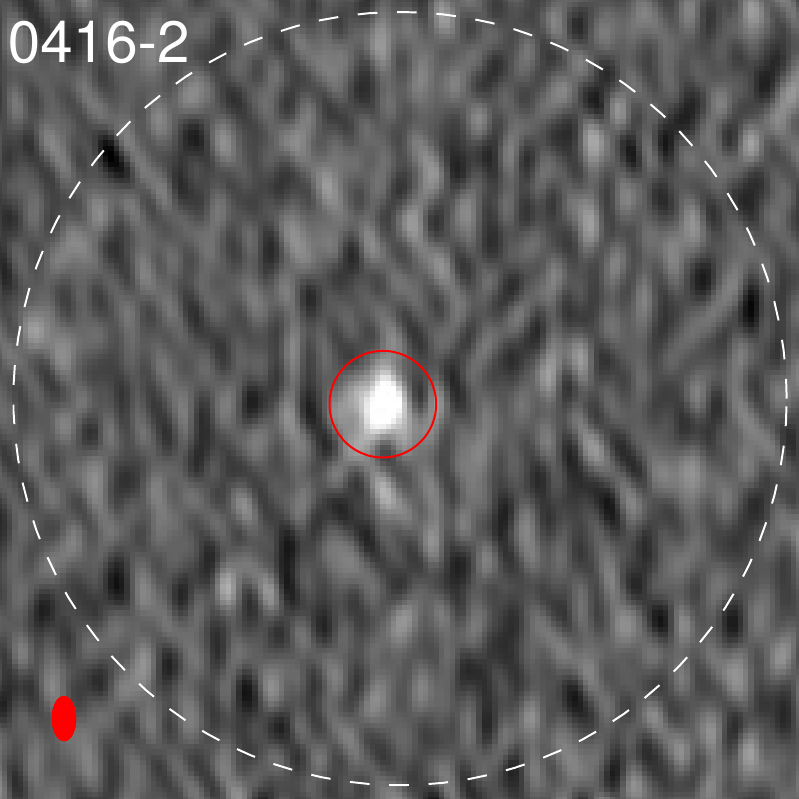}
\includegraphics[width=2.95cm]{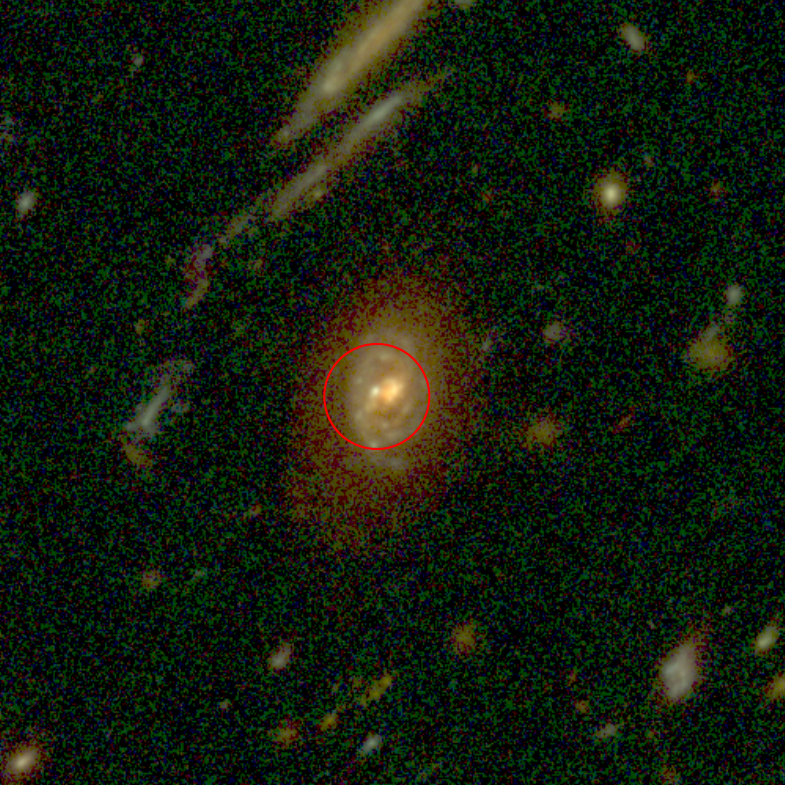} 
 \includegraphics[width=2.95cm]{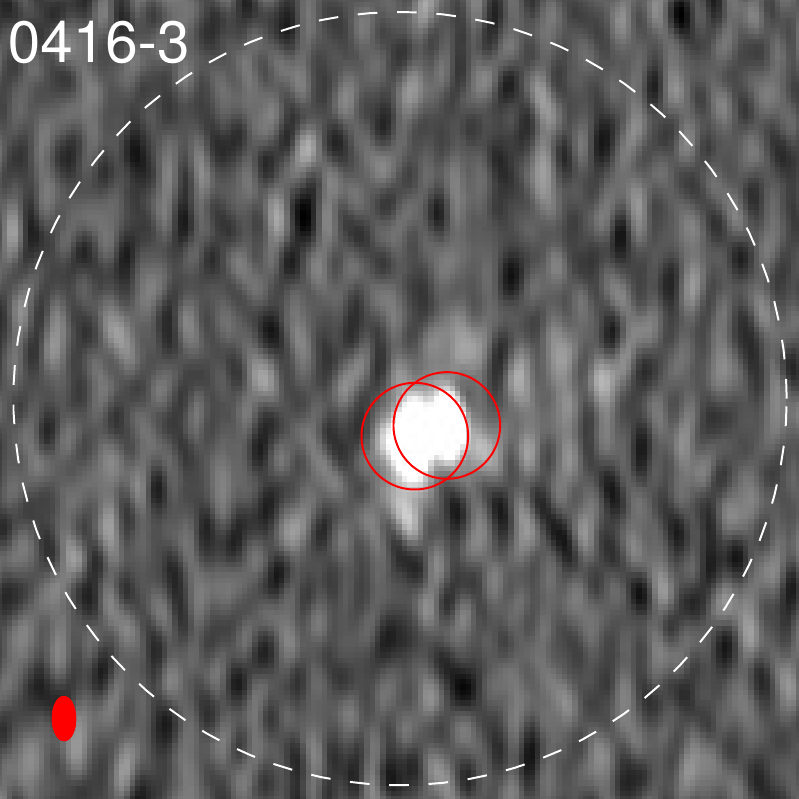}
\includegraphics[width=2.95cm]{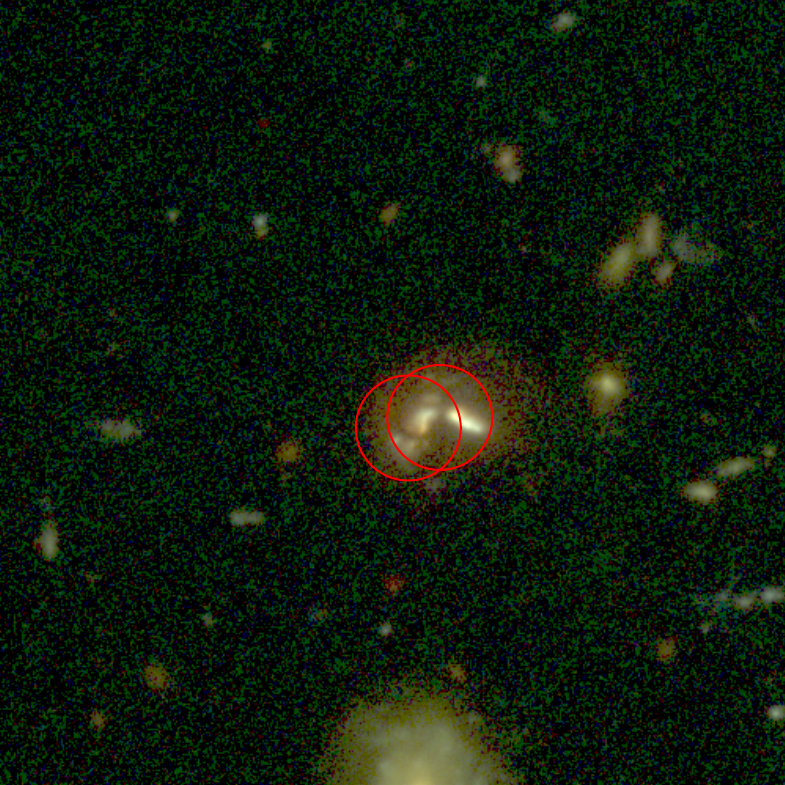} \\[-2.8mm]

 \includegraphics[width=2.95cm]{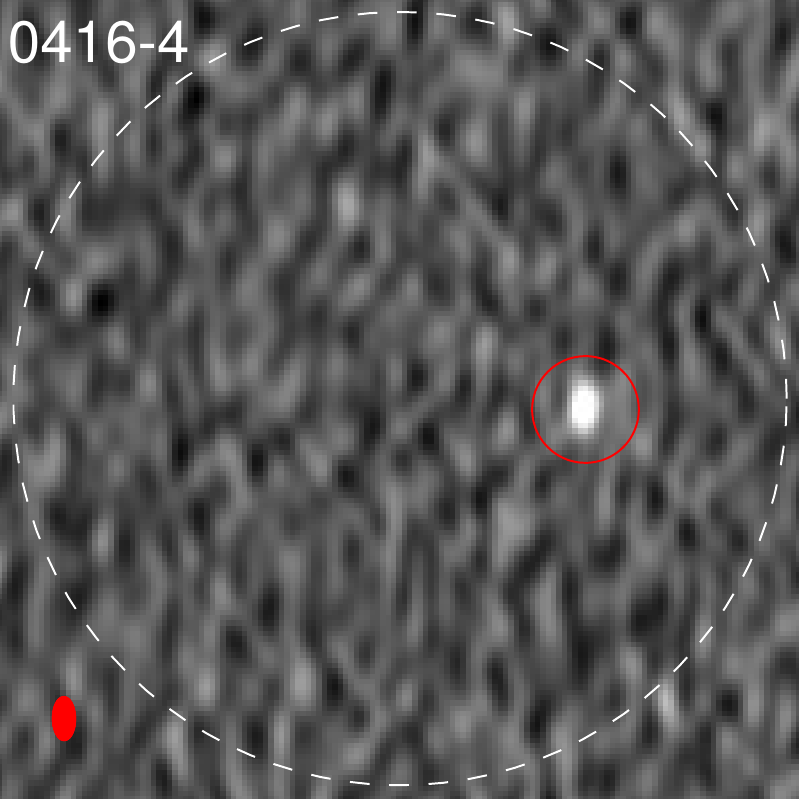}
\includegraphics[width=2.95cm]{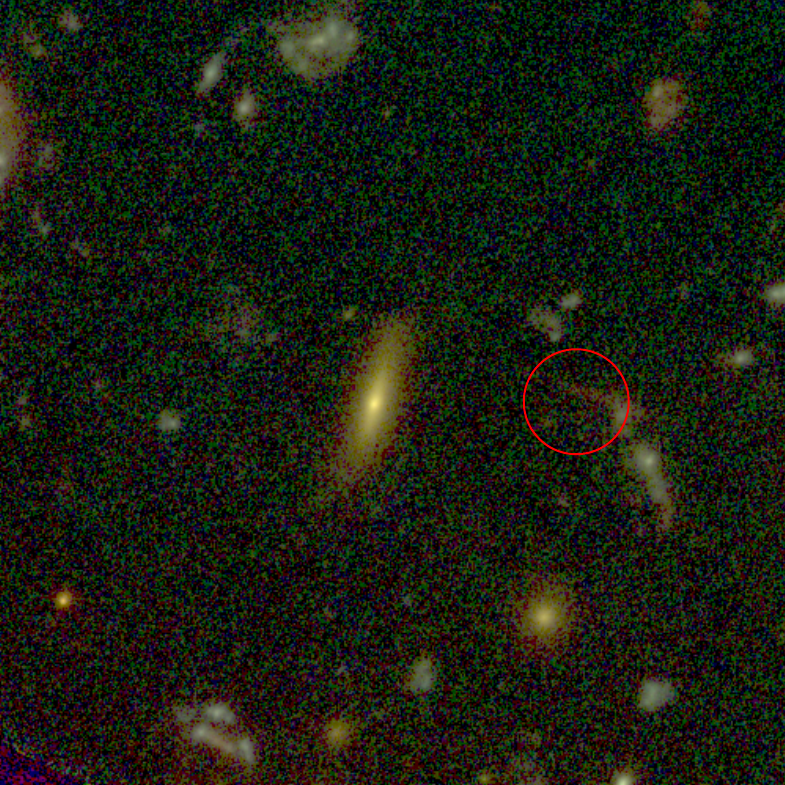}
 \includegraphics[width=2.95cm]{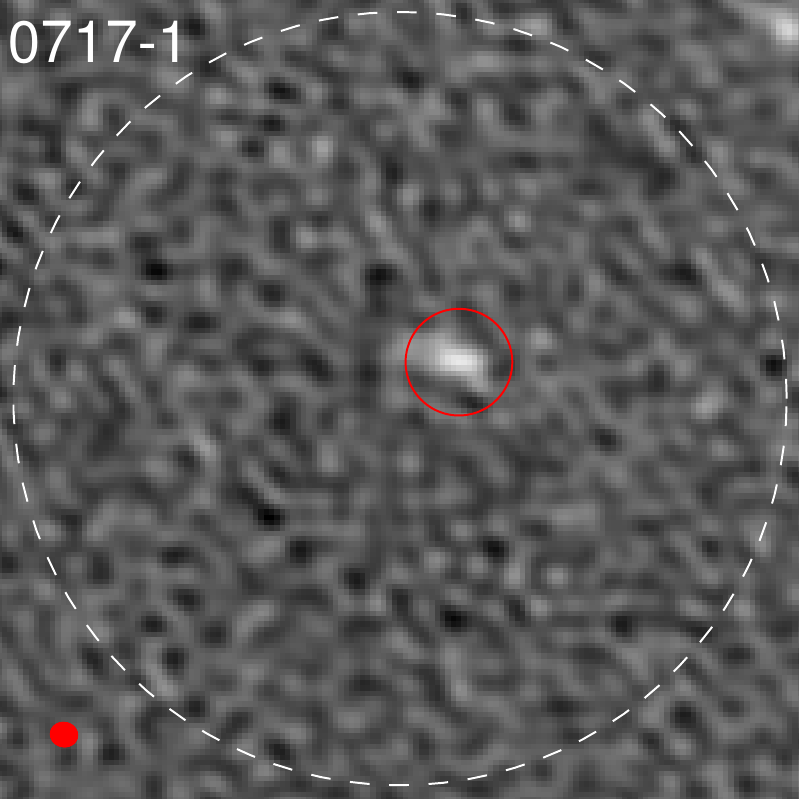}
\includegraphics[width=2.95cm]{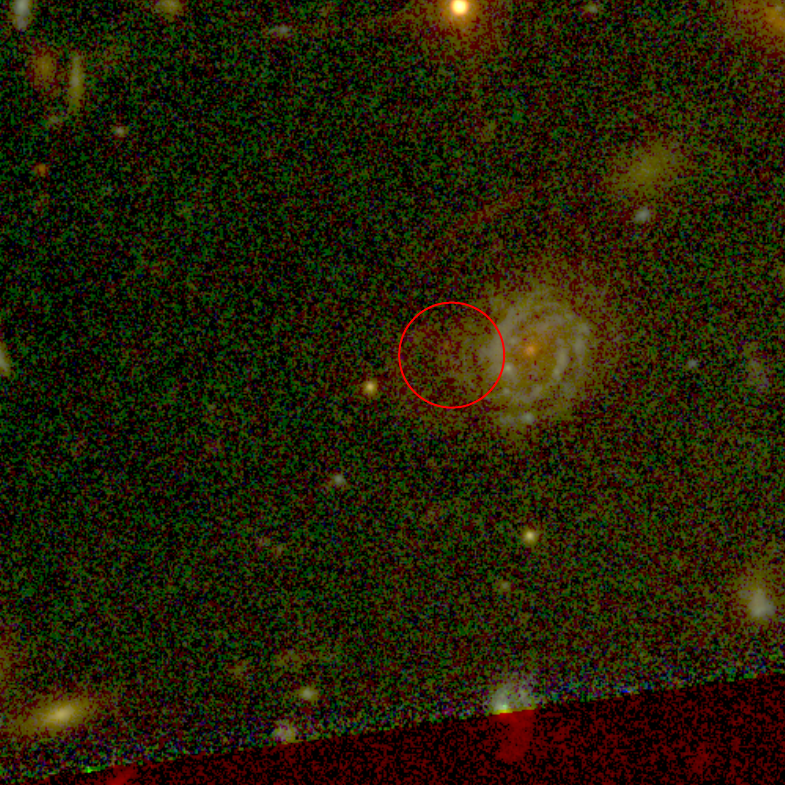} 
\includegraphics[width=2.95cm]{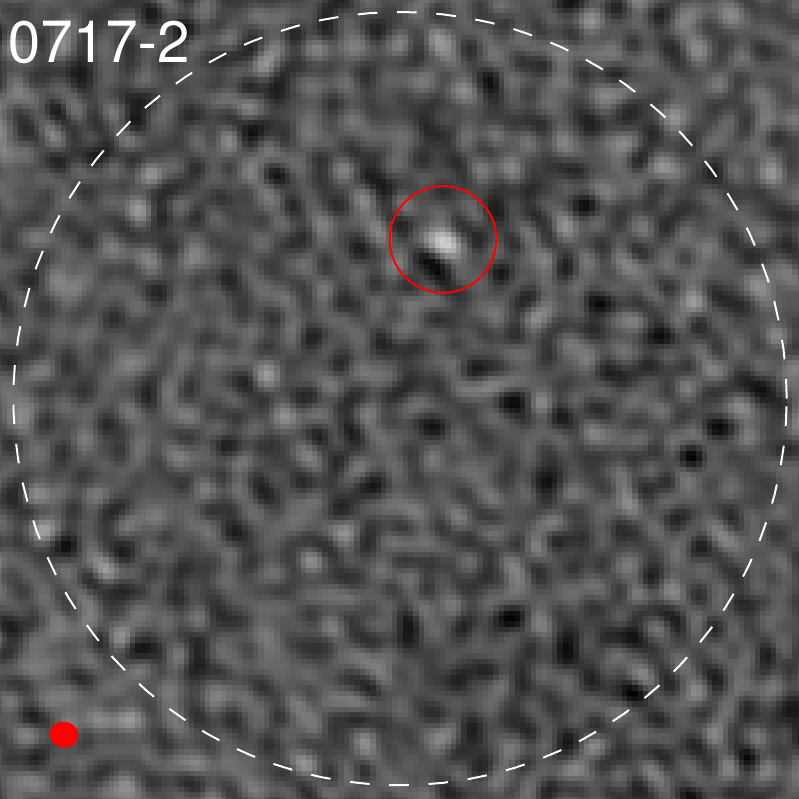}
\includegraphics[width=2.95cm]{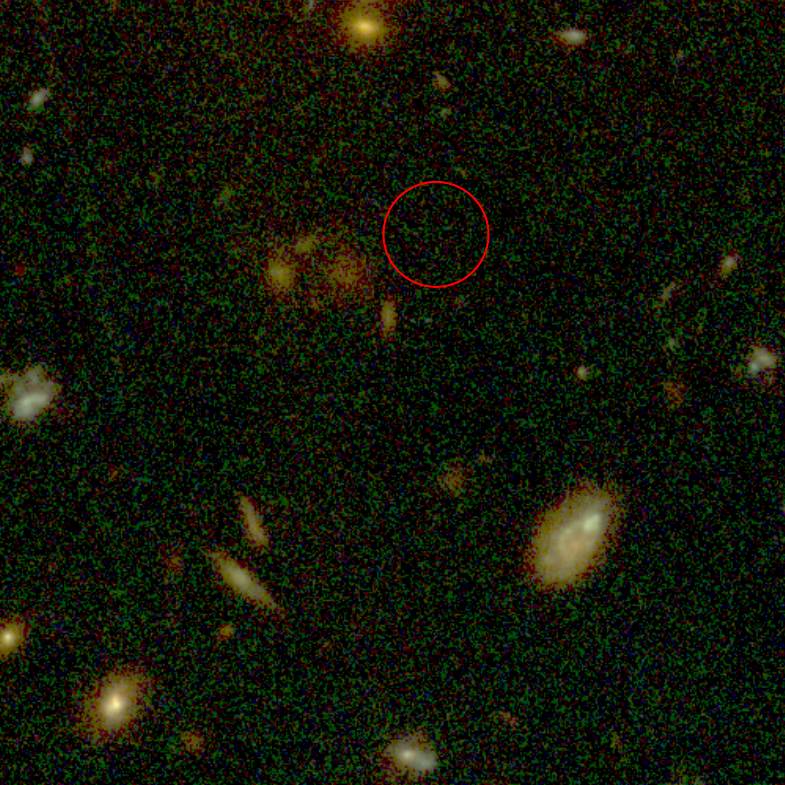} \\[-2.8mm]

\includegraphics[width=2.95cm]{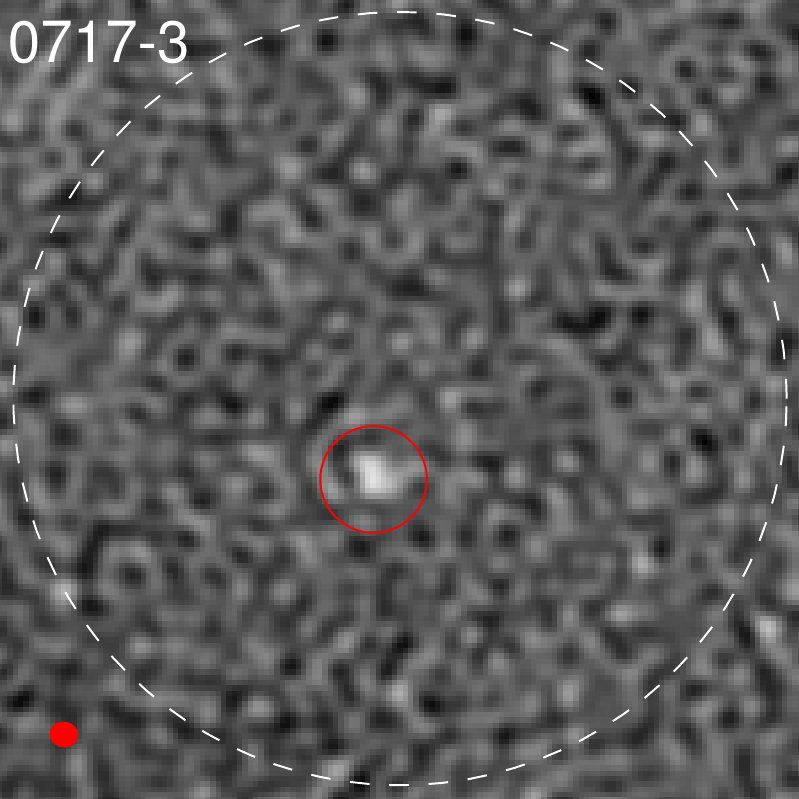}
\includegraphics[width=2.95cm]{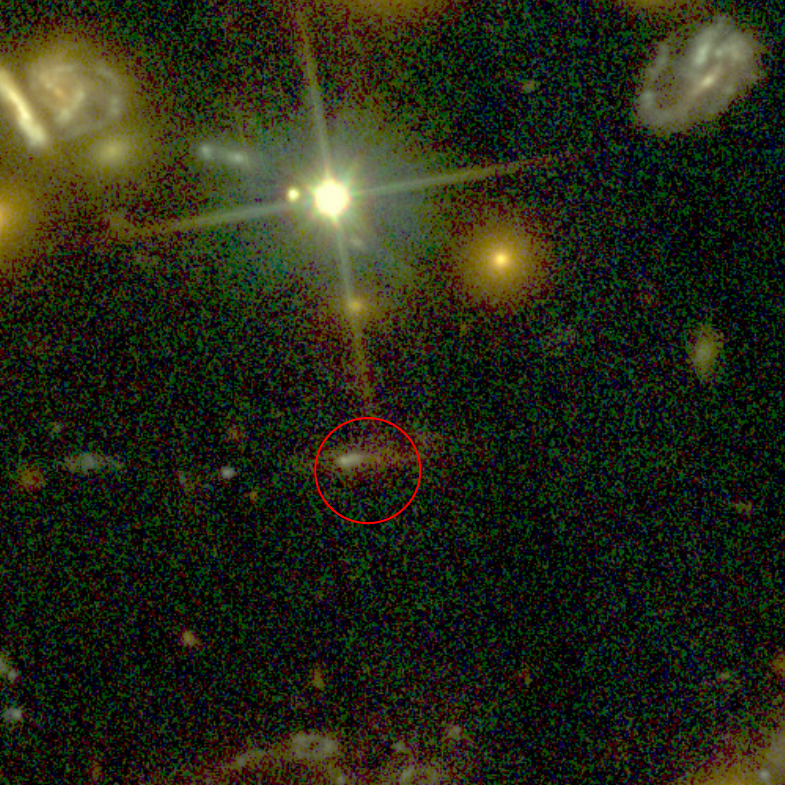}
\includegraphics[width=2.95cm]{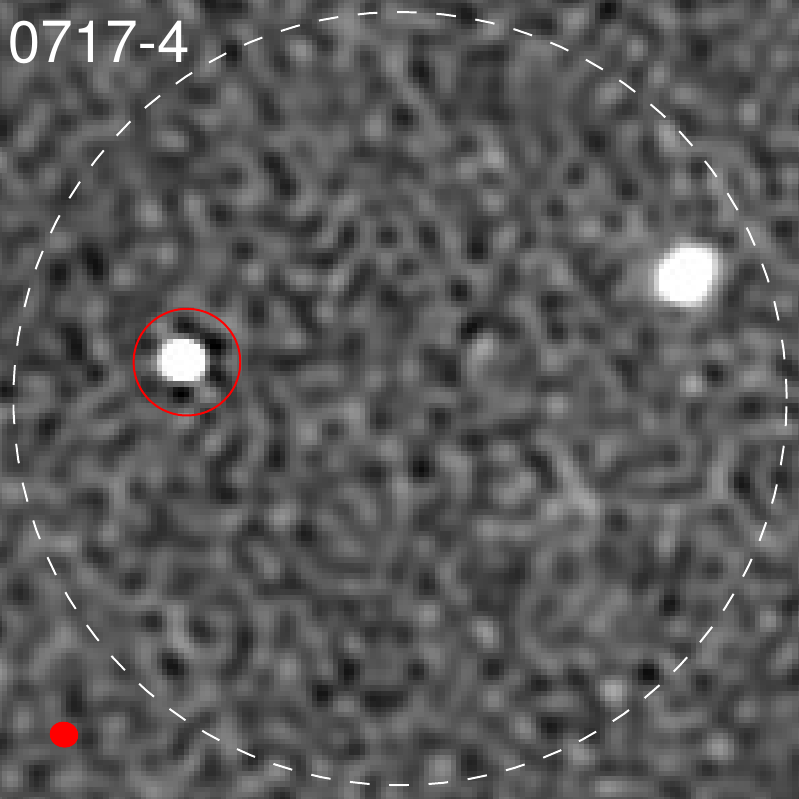}
\includegraphics[width=2.95cm]{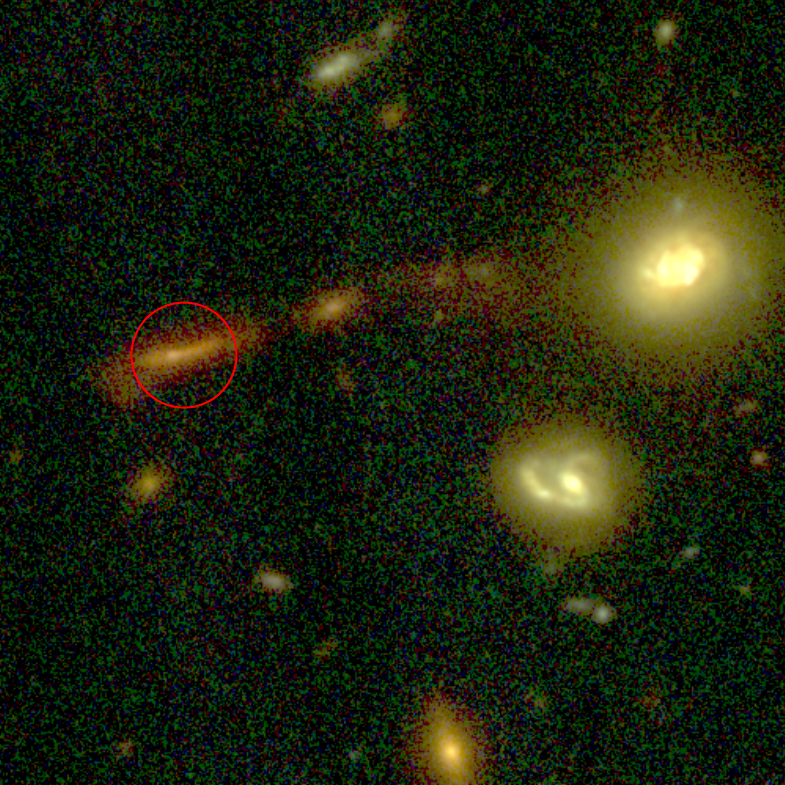}
\includegraphics[width=2.95cm]{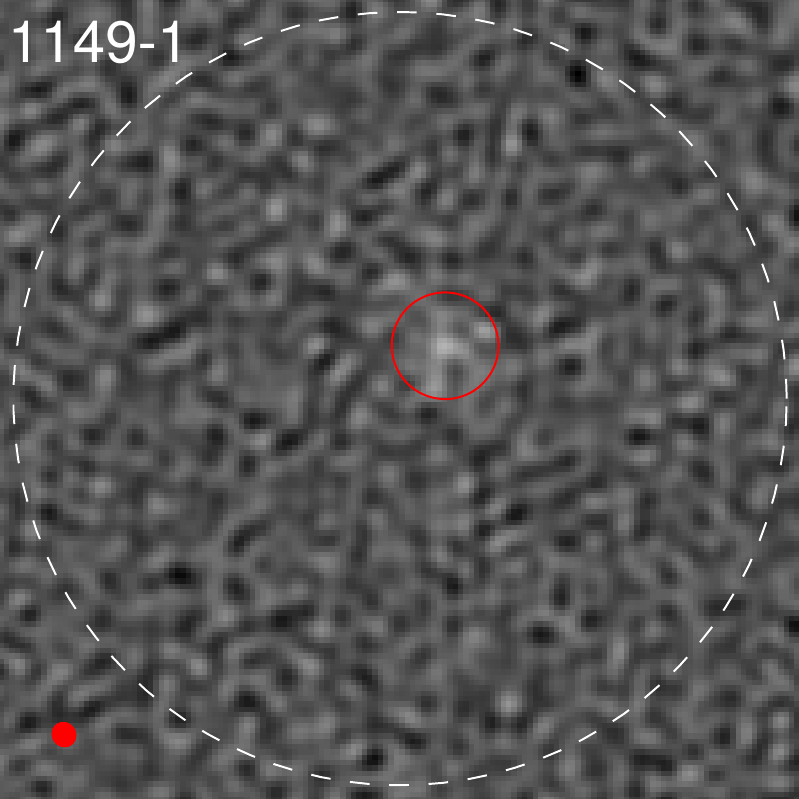}
\includegraphics[width=2.95cm]{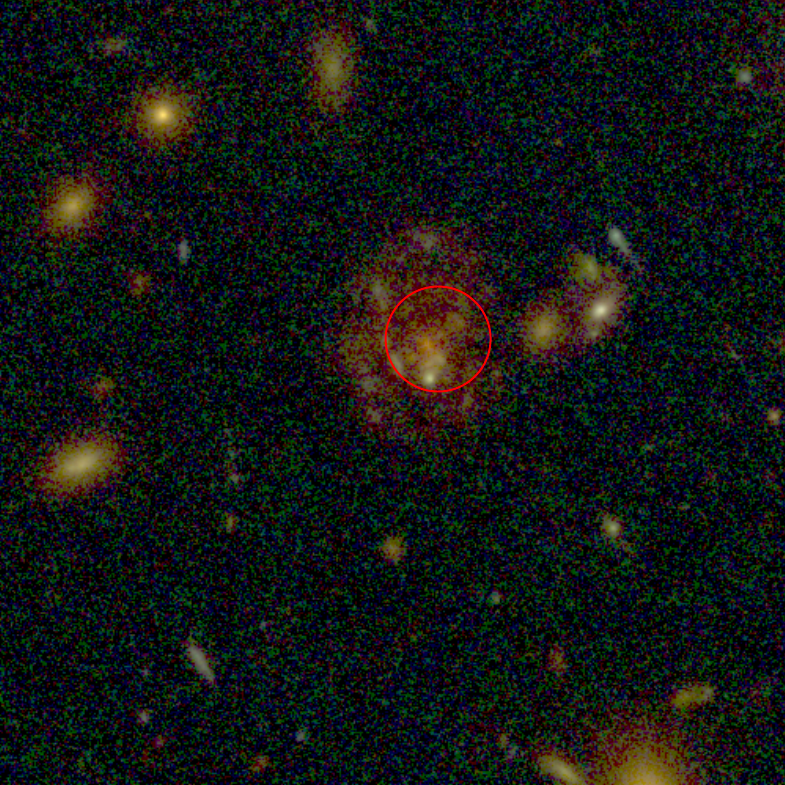}  \\[-2.8mm]

\includegraphics[width=2.95cm]{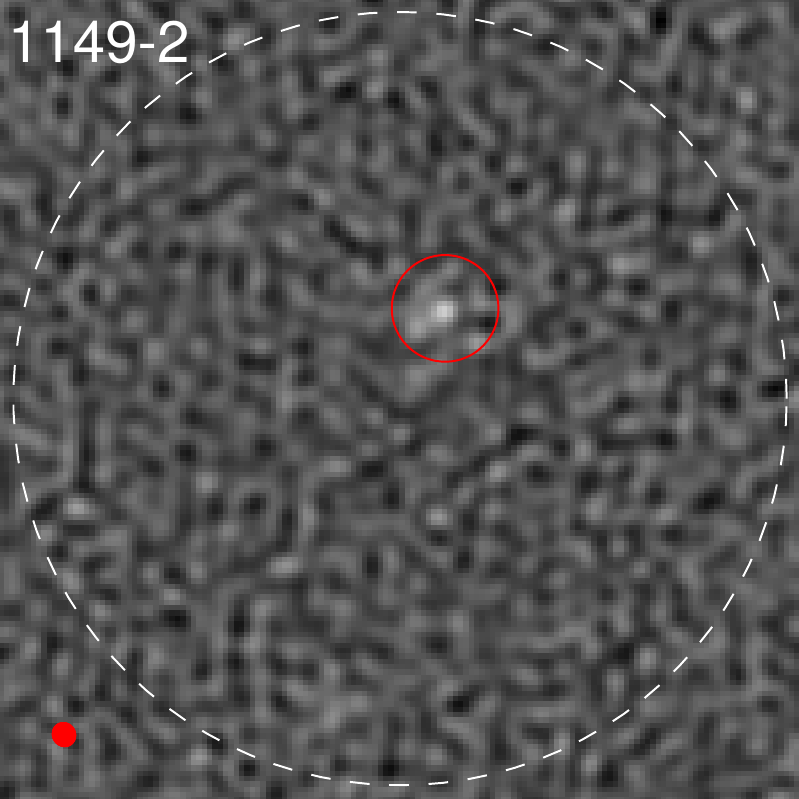}
\includegraphics[width=2.95cm]{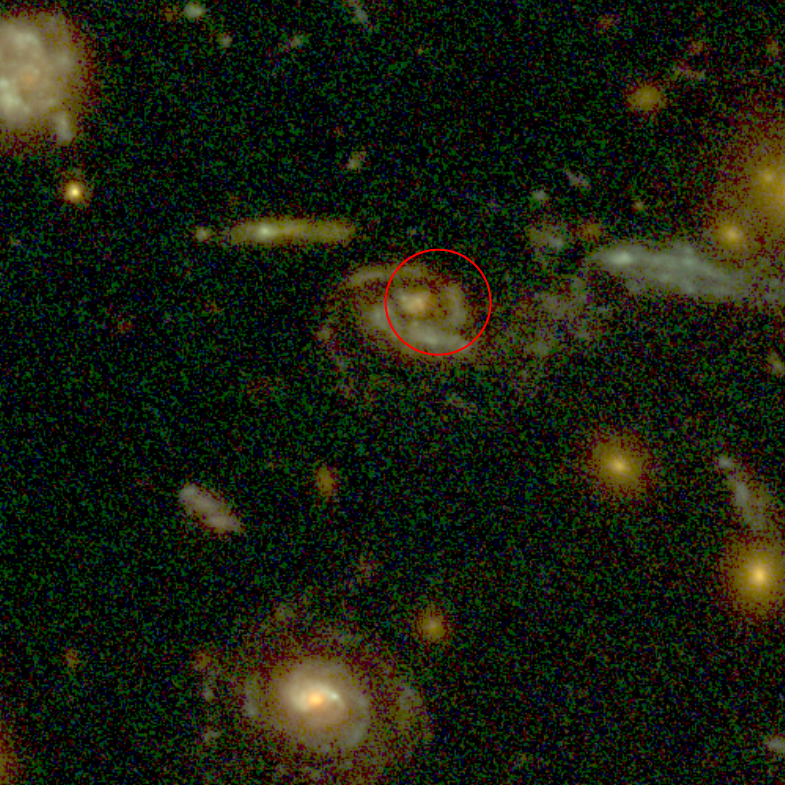}
\includegraphics[width=2.95cm]{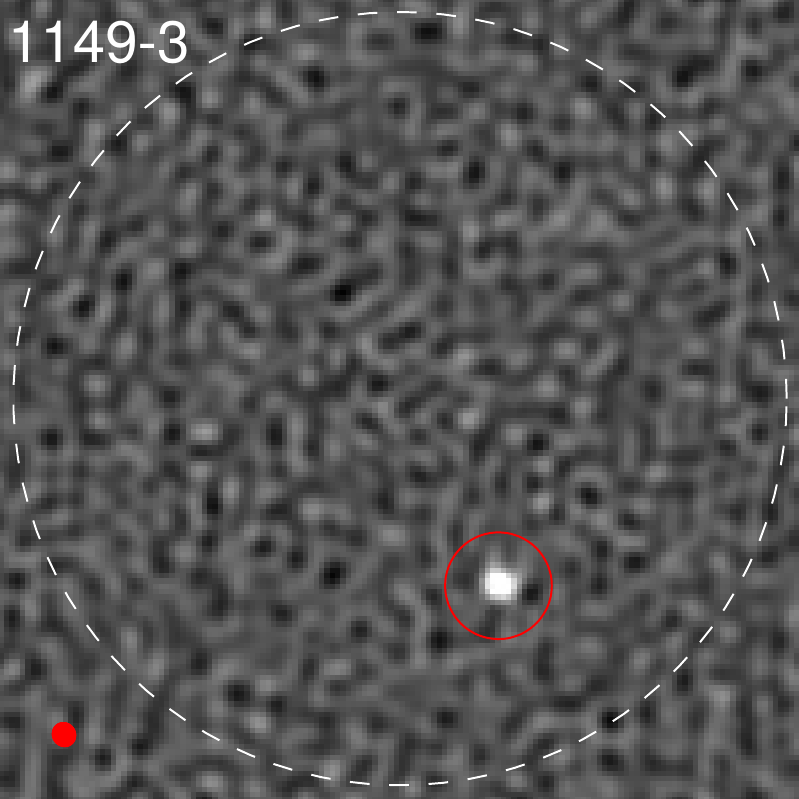}
\includegraphics[width=2.95cm]{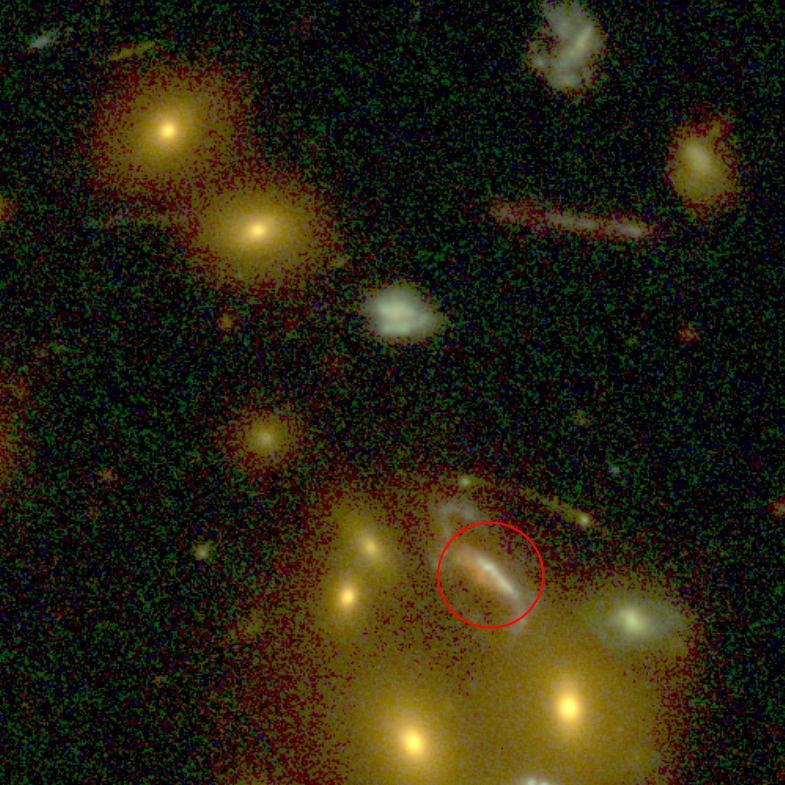}
\includegraphics[width=2.95cm]{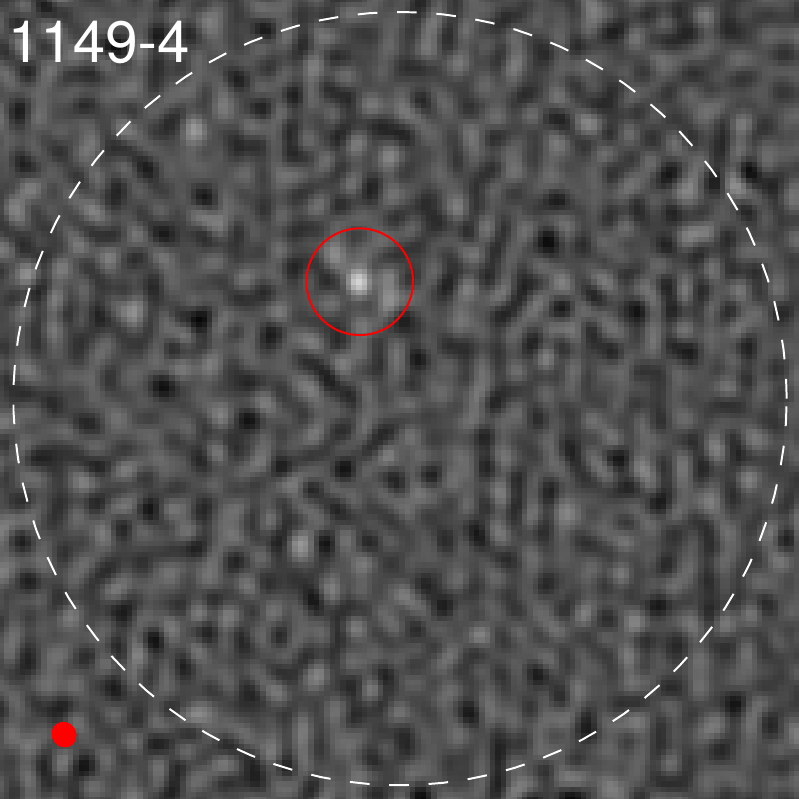}
\includegraphics[width=2.95cm]{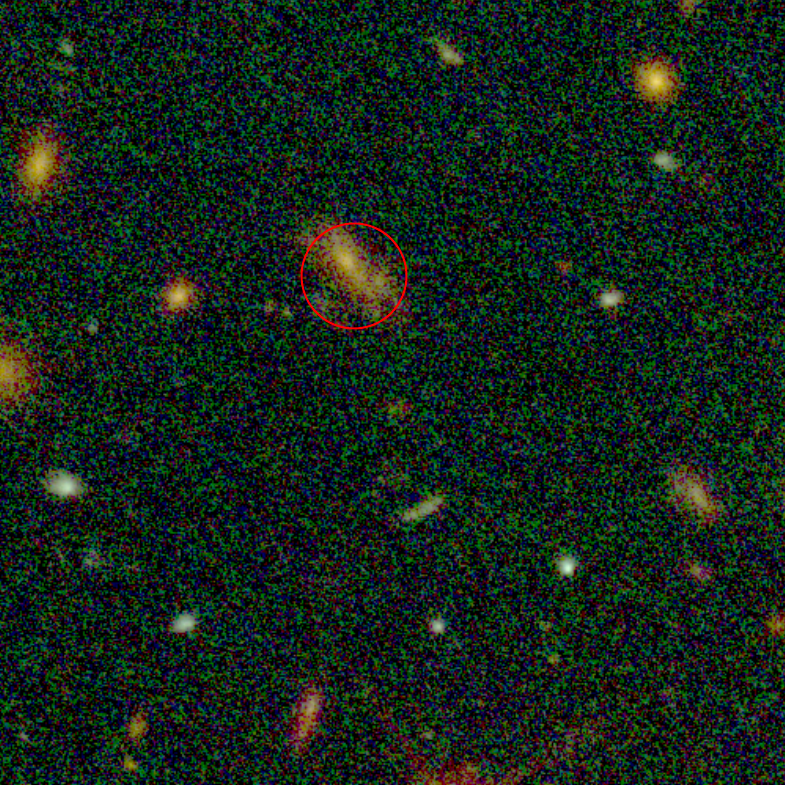}\\[-2.8mm]

\includegraphics[width=2.95cm]{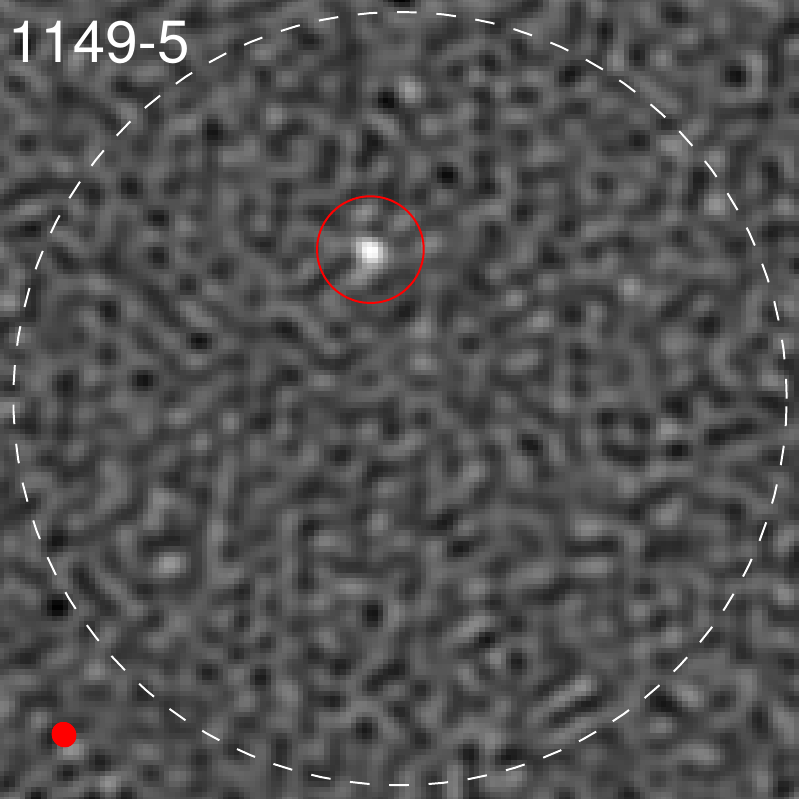}
\includegraphics[width=2.95cm]{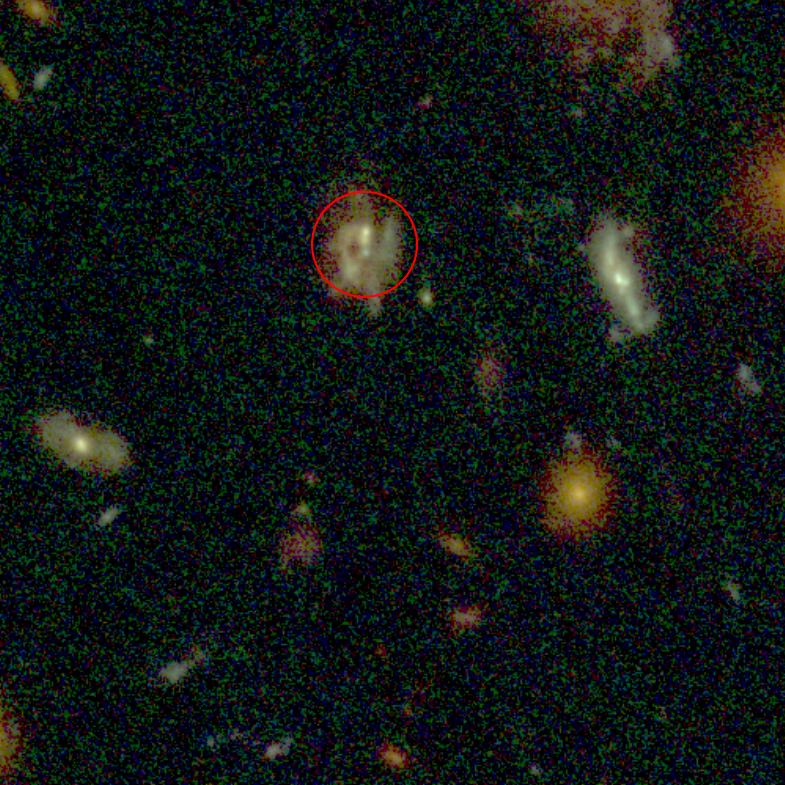} 
\includegraphics[width=2.95cm]{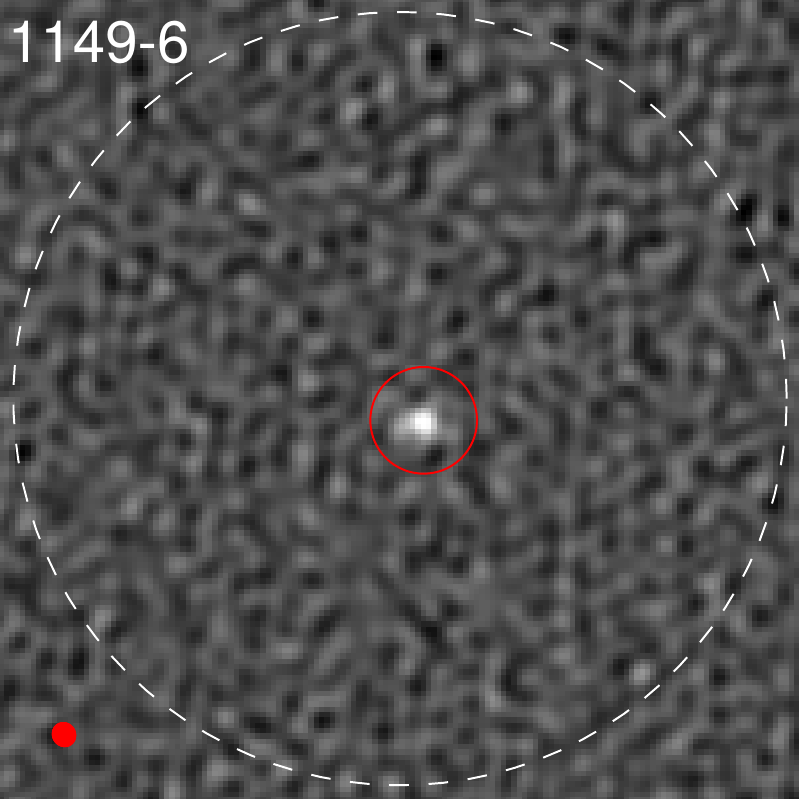}
\includegraphics[width=2.95cm]{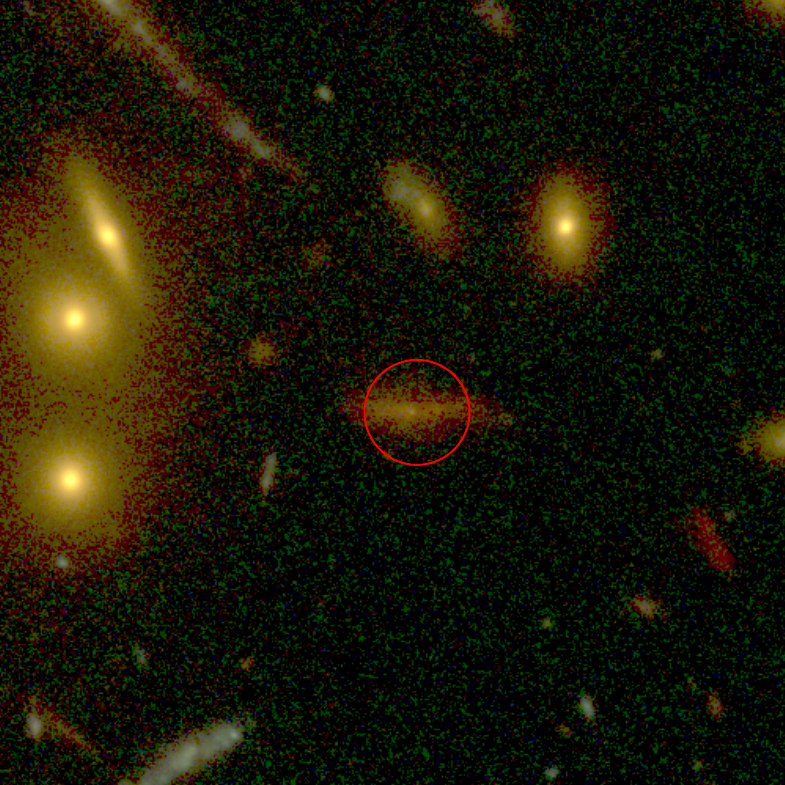} 

\caption{3 GHz identified sample of SCUBA-2 850 $\mu$m sources that are within the {\it HST} Frontier Fields coverage. For each source, we show the 3 GHz image on the left and the ACS false-color 
(F435W, F606W and F814W) image on the right. The image size is 15$'' \times$ 15$''$. In the 3 GHz images, the large dashed circles with a diameter of 14$\farcs$5 represent the JCMT beam (FWHM) at 
850 $\mu$m. The positions of the 3 GHz counterparts are indicated by the 1$''$-radius red circles in both the 3 GHz and ACS images. The ellipse at the bottom-left corner of each 3 GHz image represents 
the synthesized beam. Note that another bright radio source also locates within the SCUBA-2 beam of 0717-4. However, this source is not considered the counterpart because $p > 0.05$ and it is at $z \sim 0.3$.}

\label{fig:figure3}
\end{figure*}    

\begin{table*}
\caption{{\it Herschel} and ALMA 1.1 mm flux densities from \cite{Rawle2016A-complete-cens} and \cite{Gonzalez-Lopez2017The-ALMA-Fronti}.}
\begin{center}
\begin{tabular}{cccccccc}
\hline \hline	
 ID  &  $S_{100 \mu{\rm m}}$ & $S_{160 \mu{\rm m}}$ & $S_{250 \mu{\rm m}}$ & $S_{350 \mu{\rm m}}$ & $S_{500 \mu{\rm m}}$ & $S_{1100 \mu{\rm m}}$  \\ 
                 &   (mJy)   &  (mJy)   & (mJy)  &  (mJy) & (mJy)  & (mJy)\\
\hline
0416--1    & 5.1 $\pm$ 1.0   &  12.6 $\pm$ 2.0    & (9.7 $\pm$ 3.1)  & (6.8 $\pm$ 2.6)  & ... & 1.32 $\pm$ 0.10 \\
0416--2    &11.1 $\pm$ 1.2  &  24.5 $\pm$ 2.2    & (9.1 $\pm$ 3.5)  & (4.7 $\pm$ 2.9)       & ... & ...  \\
0416--3    & 20.1 $\pm$ 1.7 &  36.4 $\pm$ 3.2    & 31.5 $\pm$ 3.9  & 24.2 $\pm$ 3.5   &  10.5 $\pm$ 3.7  & ... \\      
0416--4    &  ...                  &   ...                       & ...                      & ...                     & ...                     & ... \\
0717--1    & 3.8 $\pm$ 0.8  &  15.2 $\pm$ 1.8     & 31.0 $\pm$ 3.8   & 34.2 $\pm$ 4.7  &  16.9 $\pm$ 3.0 & ... \\
0717--2    & ...                   &   ...                       & ...                       & ...                     & ...                     & ... \\     
0717--3    & ...                   &   ...                       & ...                       & ...                     & ...                    & ...  \\
0717--4    & ...                   &   ...                       & ...                       & ...                     & ...                    & ...  \\
1149--1    & 7.0 $\pm$ 0.9  &   16.4 $\pm$ 1.8    & 32.3 $\pm$ 3.2    & 30.3 $\pm$ 3.2  & 21.4 $\pm$ 2.8 & ...  \\
1149--2    & 3.7 $\pm$ 0.7  &   7.9 $\pm$ 1.5      & (5.9 $\pm$ 2.5)       &  ...                    & ... & 0.58 $\pm$ 0.13 \\      
1149--3    & 7.7 $\pm$ 1.0  &   18.4 $\pm$ 1.9    & 23.7 $\pm$ 3.7   &  14.2 $\pm$ 3.0  & 6.1 $\pm$ 3.9 & $<$ 0.57 \\
1149--4    & 3.2 $\pm$ 0.7  &   9.6 $\pm$ 1.4      & ...                        & ...                   & ... & ... \\
1149--5    & 5.0 $\pm$ 0.8   &  10.0 $\pm$ 1.4    & 12.5 $\pm$ 2.5    & 5.7 $\pm$ 2.8   & ... & ... \\    
1149--6    & 6.2 $\pm$ 0.9   &  11.3 $\pm$ 1.7    & 12.6 $\pm$ 2.6  &  7.4 $\pm$ 3.3    & 3.8 $\pm$ 3.4 & ... \\

\hline\hline 
\end{tabular}
\label{tab:table6}
\tablecomments{Flux densities enclosed by parentheses are flagged in the FIR fits described in Section~\ref{sec:FIR}.}  

\end{center}
\end{table*}

\begin{table*}
\caption{Dust temperatures, IR luminosities, and SFRs} 
\begin{center}
\begin{tabular}{cccccccc}
\hline \hline	
 ID   & $T$  & $L_{\rm IR}$   &SFR$_{\rm IR} $  & SFR$_{\rm 6GHz}$ & SFR$_{\rm UV}$ \\ 
           &     (K) & ($10^{11} L_{\odot}$)  & ($M_{\odot}$~yr$^{-1}$)  &  ($M_{\odot}$~yr$^{-1}$) & ($M_{\odot}$~yr$^{-1}$)  \\
\hline

0416--1   &    51.2 $\pm$ 2.0             &17.0$^{+2.3}_{-3.4}$     & 253$^{+34}_{-51}$  & \tablenotemark{*}226$^{+40}_{-71}$      & \tablenotemark{*}28.2$^{+1.2}_{-6.4}$ \\   
0416--2    &  41.2$^{+2.4}_{-2.3}$             &4.85$^{+0.57}_{-0.49}$ & 72$^{+8}_{-7}$              & 67 $\pm$ 11           & 2.3$^{+0.2}_{-0.1}$ \\             
0416--3    &  44.8$^{+2.6}_{-2.4}$              &9.76$^{+3.84}_{-2.31}$     & 146$^{+57}_{-34}$  & 281$^{+87}_{-45}$ & 5.6$^{+2.0}_{-1.0}$ \\           
0416--4    &      ...                                & ... & ... &  519$^{+1601}_{-482}$ & ...  \\   

0717--1    &    29.3 $\pm$ 1.8 &               8.34$^{+1.52}_{-1.99}$&124$^{+23}_{-30}$   & 99$^{+23}_{-31}$  & 4.3$^{+0.8}_{-1.2}$ \\   
0717--2    &    ...                               & ...  & ... & 372$^{+481}_{-225}$ & ... \\        
0717--3    &   27.9$^{+19.1}_{-8.9}$ & ... & ... & 70$^{+20}_{-24}$  & 0.8 $\pm$ 0.2 \\      
0717--4    &   17.9$^{+14.9}_{-5.8}$ & ... & ... & 112$^{+27}_{-24}$  & 0.1$^{+0.2}_{-0.1}$ \\      

1149--1     &   35.6$^{+2.6}_{-2.5}$ & 10.9$^{+2.4}_{-2.3}$     & 162$^{+36}_{-34}$  & 176$^{+75}_{-60}$ & 1.5$^{+0.7}_{-0.4}$ \\ 
1149--2    &    40.4$^{+4.8}_{-3.8}$ & 2.40$^{+0.61}_{-0.47}$ &  36$^{+9}_{-7}$     & 44$^{+14}_{-15}$   & 2.3$^{+0.5}_{-0.4}$ \\         
1149--3    &   38.2$^{+1.9}_{-1.7}$ & 2.43$^{+0.51}_{-0.71}$ & 36$^{+8}_{-11}$        & 28$^{+7}_{-8}$             & 1.2$^{+0.2}_{-0.3}$ \\   
1149--4    &   41.3$^{+5.5}_{-4.8}$ & 5.47$^{+1.25}_{-1.37}$ & 82$^{+19}_{-20}$    & 89$^{+29}_{-31}$   & 0.7 $\pm$ 0.1 \\  
1149--5    &  54.9$^{+6.0}_{-5.2}$  &  8.50$^{+1.56}_{-1.65}$      & 127$^{+23}_{-25}$ &121$^{+40}_{-28}$  & 3.2$^{+1.2}_{-0.8}$ \\      
1149--6    &   42.4$^{+4.2}_{-4.0}$ &  3.65$^{+1.01}_{-0.72}$ & 55$^{+15}_{-11}$           & 63$^{+21}_{-16}$   & 0.1 $\pm$ 0.1 \\    
\hline \hline
\end{tabular}  
\label{tab:table7}
\tablecomments{The uncertainties of dust temperatures include photometric errors and the uncertainties of redshifts. The uncertainties of IR luminosities and SFRs include photometric errors as well 
as the uncertainties of redshifts and lensing magnifications. \tablenotemark{*}Because 0416-1 is classified as an AGN in the GLASS survey \citep{Schmidt2014Through-the-Loo,Treu2015The-Grism-Lens-}, it is 
unclear if the radio and UV emission is dominated by the AGN or star formation. We caution that the radio and UV SFRs for this source can only be considered as upper limits. }  
\end{center}
\end{table*}

\section{Properties of Radio-detected Submillimeter Sources}\label{sec:properties}

\subsection{Dust Temperatures and IR SFRs}\label{sec:FIR}

We cross-match our final sample of SCUBA-2 sources with the source catalog from the {\it Herschel Lensing Survey} \citep{Rawle2016A-complete-cens}. Using the 3 GHz positions as prior, 10 of the 14 sources 
have a nearest {\it Herschel} detection with an offset of $<$ 1$''$. On the other hand, the remaining four sources have no {\it Herschel} counterpart even when a 10$''$ matching radius is used. For the 10 
sources with {\it Herschel} detections, the optical counterparts we identified completely agree with the optical counterparts Rawle et al. found. Additionally, 0416-1 and 1149-2 are detected by the Atacama 
Large Millimeter/submillimeter Array (ALMA) 1.1 mm imaging of \cite{Gonzalez-Lopez2017The-ALMA-Fronti} as MACSJ0416-ID01 and MACSJ1149-ID01, respectively. We also estimate the 5$\,\sigma$
limit of 1.1 mm flux density for 1149-3, which is the only other source covered by the ALMA maps. The observed {\it Herschel} and ALMA 1.1 mm flux densities are shown in Table~\ref{tab:table6}. 

For four sources, we found that the {\it Herschel}/SPIRE photometry does not agree well with our SCUBA-2 flux densities. Compared with the SCUBA-2 photometry, the SPIRE SEDs of 
these sources turn over at shorter wavelengths. This inconsistency might be caused by the PSF-fitting procedure and/or deblending photometry performed by 
\cite{Rawle2016A-complete-cens}. We therefore decided to flag the SPIRE flux densities that are a factor of two lower than the best-fit modified blackbody models for 
the SEDs constructed from the {\it Herschel}/PACS, SCUBA-2, and ALMA (if available) photometry. These flags are done for 0416-1, 0416-2, and 1149-2, which are presented in Table~\ref{tab:table6} and Figure~4.

For the sources with spectroscopic or photometric redshifts, we measure their dust temperatures by fitting a modified blackbody model with $\beta = 1.5$ (e.g., \citealt{Chapman2005A-Redshift-Surv,Kovacs2006SHARC-2-350-mum,Pope2006The-Hubble-Deep}) to the FIR flux densities. The resulting median dust temperature is 41.2$^{+1.8}_{-2.0}$ K. We therefore use $\beta = 1.5$ and $T = 41.2$ K to estimate the redshifts for the remaining two sources (0416-4 and 0717-2) that have no redshift measurements. Note that only 450 and 850 $\mu$m flux densities are available for these two sources as well as 0717-3 and 0717-4 because they are not detected in {\it Herschel Lensing Survey}. As a consequence, what we do is simply match the models and the 450 $\mu$m-to-850 $\mu$m flux ratios instead of least chi-squared fitting. 

We fit the templates of \cite{Rieke2009Determining-Sta} to the FIR flux densities for the 10 sources that are detected by {\it Herschel} to derive their IR luminosities 
($L_{\rm IR}$; $\lambda_{\rm rest}$ = 8-1000~$\mu$m). $L_{\rm IR}$ is then converted to SFR via the theoretical relation in \cite{Murphy2011Calibrating-Ext}. Both the $L_{\rm IR}$ and SFRs are corrected 
for the lensing magnifications based on our redshifts and the lens models. We do not compute the IR SFRs of 0416-4, 0717-2, 0717-3, and 0717-4 since they are only detected at 450 and 850 $\mu$m. We show these fits in Figure~4. All the derived quantities are listed in Figure~4 (without uncertainties) and Table~\ref{tab:table7}.

\begin{figure*}
\begin{center}    

\includegraphics[width=16.8cm]{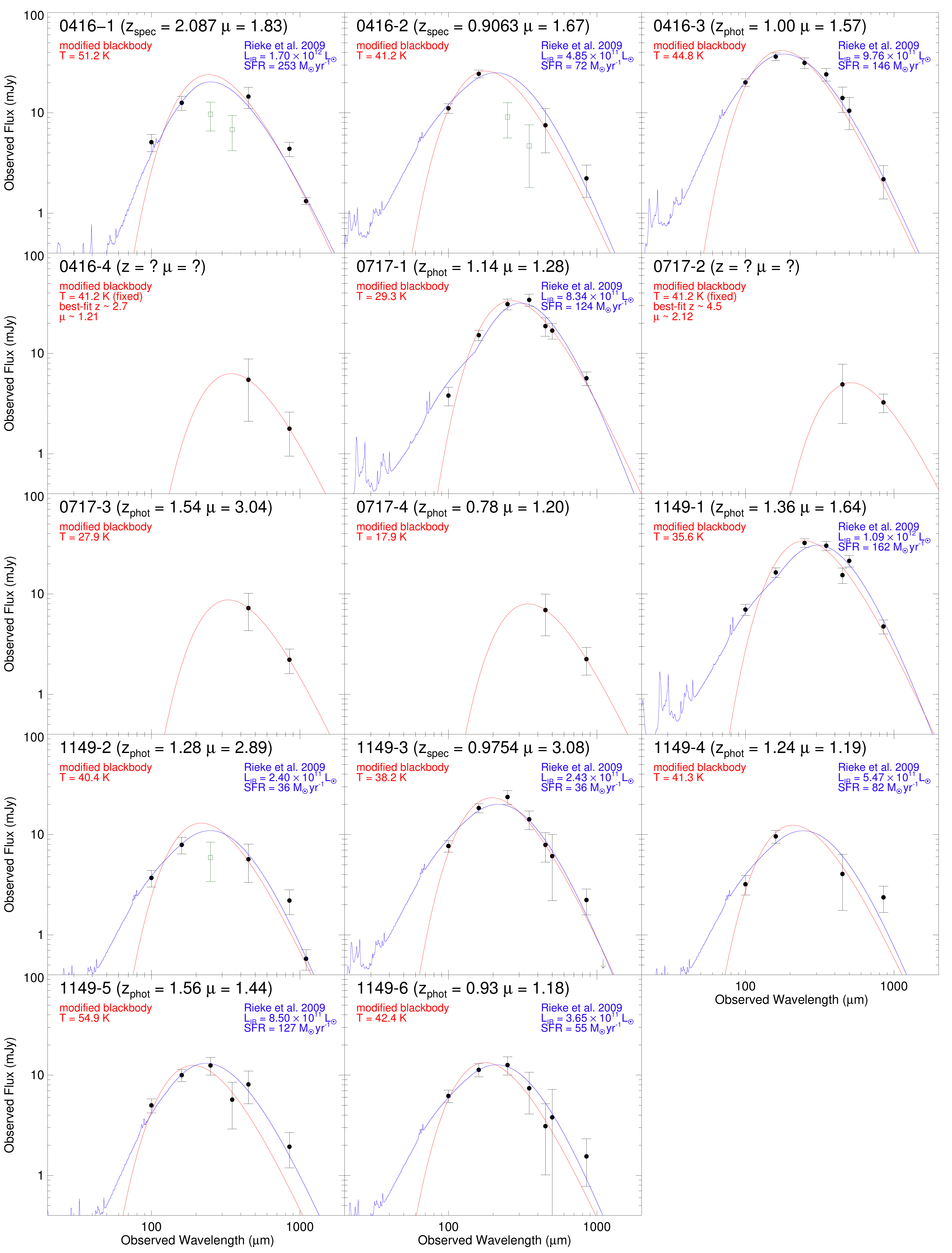}

\caption{FIR SEDs of our 3 GHz identified sample of SCUBA-2 850 $\mu$m sources that illustrate the observed photometry and fits. The flux densities 
are from {\it Herschel}/PACS (100 and 160 $\mu$m), {\it Herschel}/SPIRE (250, 350, and 500 $\mu$m), SCUBA-2 (450 and 850 $\mu$m), and ALMA (1.1mm). The {\it Herschel} and ALMA 
flux densities are from \cite{Rawle2016A-complete-cens} and \cite{Gonzalez-Lopez2017The-ALMA-Fronti}, respectively. The SPIRE flux densities shown as green squares are 
flagged in the fits. In each panel, we plot the best-fit modified blackbody (red line) and \cite{Rieke2009Determining-Sta} template (blue line).\footnote{Note that for 0416-4, 0717-2, 0717-3, and 0717-4, only the SCUBA-2 flux densities are available. Therefore, the plotted red lines are just the modified blackbody models that match the 450 $\mu$m-to-850 $\mu$m flux ratios, and are not from least chi-squared fitting.} Derived quantities from the two models are shown in matching colors, including IR luminosities and SFRs, which take the magnifications ($\mu$) into account. 12 of the 14 sources have spectroscopic ($z_{\rm spec}$) or photometric ($z_{\rm phot}$) redshifts. For the two sources without redshifts, we use a modified blackbody with the median dust temperature from the other 11 sources (41.2 K) to convert their 450 $\mu$m-to-850 $\mu$m flux ratios to redshifts and therefore magnifications. The uncertainties of all the quantities are listed in Table~\ref{tab:table7}.}

\end{center}
\label{fig:figure4}
\end{figure*}

\subsection{Radio SFRs}\label{sec:radio}

Following \cite{Murphy2011Calibrating-Ext,Murphy2012The-Star-Format}, we compute the radio SFRs of our sources using the relation
\begin{equation}
\label{eq-sfrrad}
\begin{split}
\left(\frac{\rm SFR_{\nu}}{M_{\sun}\,{\rm yr^{-1}}}\right) &= 10^{-27}  
\left[2.18 \left(\frac{T_{\rm e}}{10^{4}\,{\rm K}}\right)^{0.45} \left(\frac{\nu}{\rm GHz}\right)^{-0.1}\right. + \\
&\left.15.1 \left(\frac{\nu}{\rm GHz}\right)^{-\alpha^{\rm NT}}\right]^{-1} \left(\frac{L_{\nu}}{\rm erg~s^{-1}~Hz^{-1}}\right)
\end{split}
\end{equation}
where we assume an electron temperature of $T_{\rm e} =10^{4}$\,K, and a constant non-thermal radio spectral index of $\alpha^{\rm NT} = 0.85$, which is the average non-thermal 
spectral index found among the 10 star-forming regions in NGC\,6946 studied by \cite{Murphy2011Calibrating-Ext}. Since our sources are all detected at 3 GHz and have a median redshift close 
to one, we decide to compute the rest-frame 6 GHz SFRs. In order to $K$-correct an observed radio flux density to rest-frame 6 GHz, we need the radio spectral index, $\alpha$, which relates the radio
flux density with frequency via a power law $S_{\nu} \propto \nu^{-\alpha}$. We can then calculate the rest-frame 6 GHz radio luminosities using
\begin{equation}
\label{eq:K-corr}
L_{\nu_{\rm rest}} ({\rm 6~GHz}) = 4\pi d_{L}^{2} S_{\nu_{\rm obs}} ({\rm 3~GHz})(1+z)^{\alpha-1}  \times 2^{- \alpha}
\end{equation}
where $d_L$ is the luminosity distance. This calculation includes a bandwidth compression term of $(1 + z)^{-1}$ and a color term of $(\frac{2}{1+z})^{-\alpha}$. For the six radio sources 
that are detected at both bands, we directly compute their spectral indices using the flux densities measured from the 3 GHz images and the convolved 6 GHz images with the Gaussian fitting procedure. We obtain an average of 0.76 $\pm$ 0.12 
from these six sources, which is consistent with the values in the literature (e.g., \citealt{Ibar2009Deep-multi-freq,Ivison2010BLAST:-the-far-,Ivison2010The-far-infrare}). We assume 
this value for the other sources. The resulting radio SFRs are tabulated in Table~\ref{tab:table7}.

\subsection{UV SFRs}

While the radio and IR SFRs represent the total and dust-obscured SFRs, respectively, the unobscured contributions from the (observed) UV emission should be accounted as well, given that
most of our sources are detected in the optical images. We use rest-frame 2271 $\rm \AA$ ({\it GALEX} NUV band) flux densities and the conversion in \cite{Murphy2012The-Star-Format} to
compute the UV SFRs of our sources without extinction correction. We interpolate the {\it HST} photometry to obtain rest-frame 2271 $\rm \AA$ flux densities and then 
compute $L_{\rm NUV}$ and UV SFRs. These UV SFRs are also tabulated in Table~\ref{tab:table7}, along with radio and IR SFRs. In Figure~5, we compare 
the radio SFRs with the IR+UV SFRs for the 10 sources that have IR SFR measurements. The UV SFRs are mostly too small to significantly affect the comparison except for 0416-1.

\begin{figure}
\begin{center}    
\includegraphics[width=8cm]{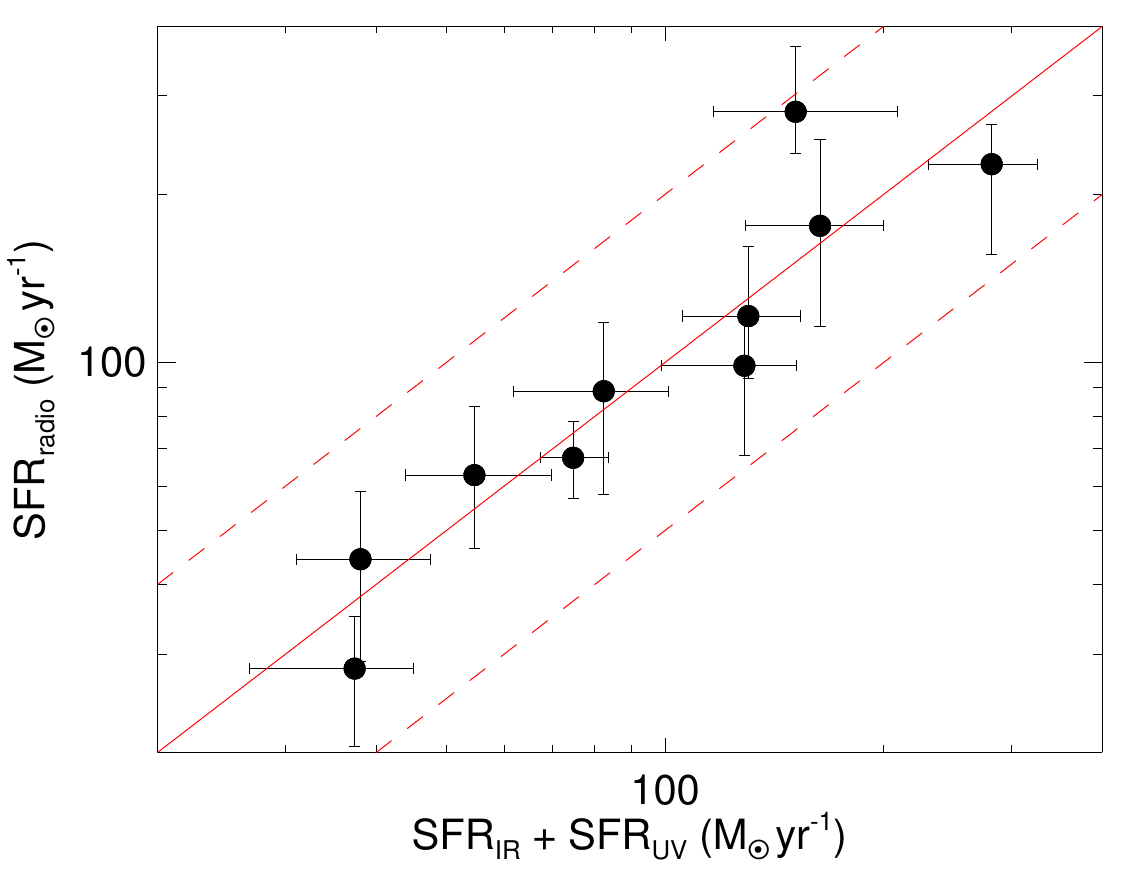}
\caption{Comparison between the radio SFRs and the IR+UV SFRs for the 10 sources that have IR SFR measurements. These SFRs are corrected for the lensing magnifications. The red solid line is the one-to-one relation. All of the 10 sources are within a multiplicative factor of two about the one-to-one line (the two red dashed lines).}
\end{center}
\label{fig:figure5}
\end{figure}

\subsection{850 $\mu$m Flux Density to SFR Conversion}\label{sec:850SFR}

The observed (but de-lensed) 850 $\mu$m flux density of an SMG should work as a proxy of the IR luminosity and IR SFR independent of redshift since the strong negative $K$-correction and the effect of distance 
almost exactly cancel out  (e.g., \citealt{Blain1993Submillimetre-C,Blain2002Submillimeter-g}). \cite{Barger2014Is-There-a-Maxi} and \cite{Cowie2017A-Submillimeter} have both measured the mean conversion 
between observed 850 $\mu$m flux density and IR SFR from their samples, with a multiplicative range over the individual values of two in each direction about the mean. Here we perform the same exercise, comparing our observed 850 $\mu$m flux densities and IR SFRs for the 10 sources that have IR SFR measurements in Figure~6. The median conversion of these sources is ${\rm SFR}_{\rm IR} (M_{\odot}~\rm{yr}^{-1})$ $= 54 \times S_{850\mu{\rm m}}$ (mJy), which is more than a factor of two smaller than the conversion (143) found in \cite{Cowie2017A-Submillimeter}. However, the median redshifts of the two samples are $z = 1.24$ (this work) and $z = 2.28$ (Cowie et al.). Therefore, it is not surprising to see different properties between the two samples.

The large difference of $\left \langle  {\rm SFR}_{\rm{IR}} / S_{850\mu{\rm m}} \right \rangle$ between the two studies is caused by the different SEDs of the two samples. If we only consider the cold dust emission at FIR wavelengths, the mean SED of our sample is close to a modified blackbody with $\beta = 1.5$ and $T = 41.2$ K. The sources in \cite{Cowie2017A-Submillimeter}, on the other hand, are generally well described by an optically thin modified blackbody ($S_{\nu} \propto \nu^{\beta} B_{\nu}(T)$) with $\beta = 1.25$ and $T = 43$ K. If the modified blackbody model in this work is used, the resulting dust temperature would be $\sim$ 50 K. The main difference between these two samples is therefore in the dust temperature (or equivalently, the peak wavelength $\lambda_{\rm peak}$), and they make more than a factor of two difference in the 
contribution to IR luminosity at the same redshift.

The variation of ${\rm SFR}_{\rm{IR}} / S_{850\mu{\rm m}}$ among our sources is also a result of different dust temperatures. The three higher outliers in Figure~6 are 0416-1, 0416-3, and 1149-5, which 
have higher dust temperatures than the rest of the sample. In contrast, the only lower outlier in Figure~6 is 0717-1, which has the lowest dust temperature among the 10 sources that have IR SFR measurements. Note that different contributions from 
the emission at shorter wavelengths can be another cause of the different ${\rm SFR}_{\rm{IR}} / S_{850\mu{\rm m}}$. This result shows that our sample of low-redshift faint SMGs has lower dust temperatures (longer $\lambda_{\rm peak}$) than those of the bright SMGs, in agreement with other studies (e.g., \citealt{Casey2012A-Population-of,U2012Spectral-Energy,Lee2013Multi-wavelengt,Symeonidis2013}).

\begin{figure}
\begin{center}    
\includegraphics[width=9cm]{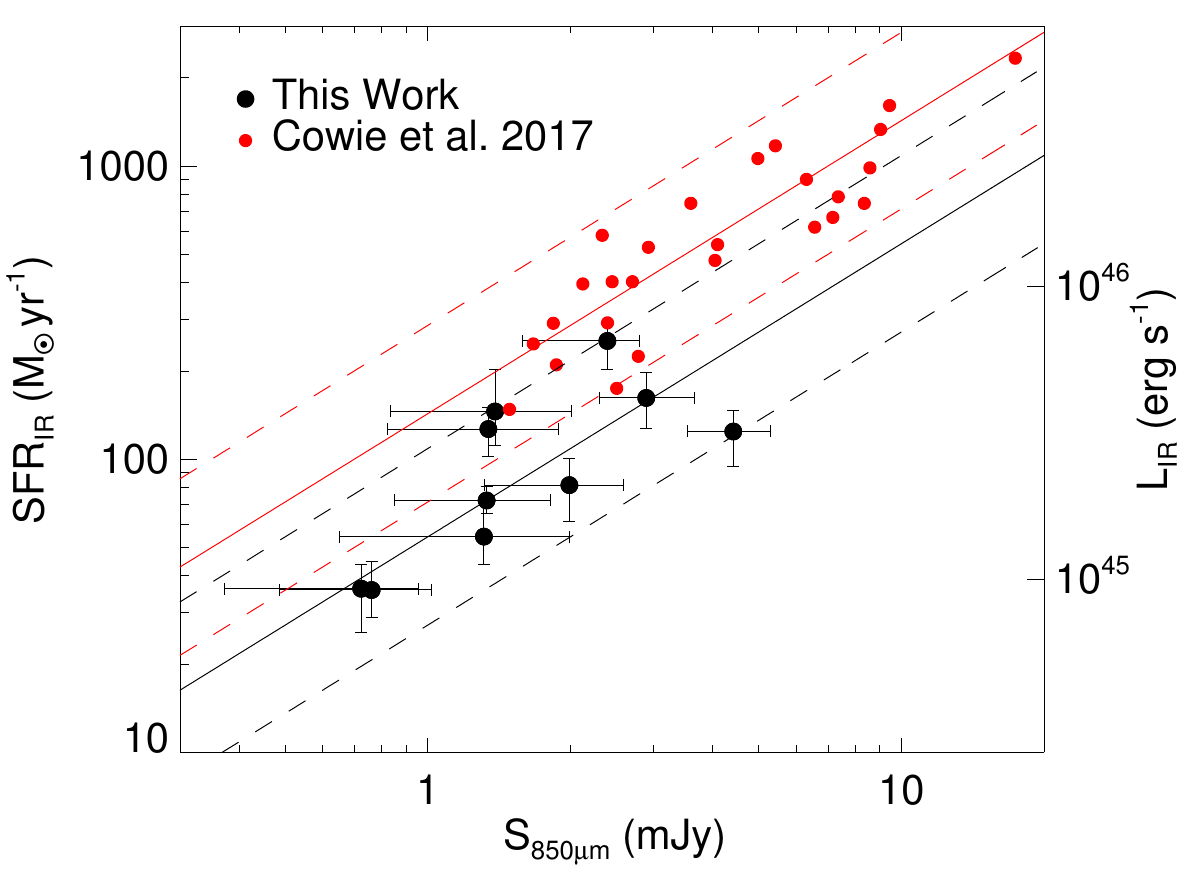}
\caption{Comparison between the 850 $\mu$m flux densities and IR SFRs/luminosities for the 10 sources that have IR SFR measurements. Both of these quantities are corrected for the lensing magnifications. 
The black solid line is the median conversion of these 10 sources, ${\rm SFR_{IR}} (M_{\odot}~\rm{yr}^{-1})$ $= 54 \times S_{850\mu{\rm m}}$ (mJy), and the black dashed lines correspond to a multiplicative 
factor of two about the black solid line. The red circles are the spectroscopic sample of SMGs from \cite{Cowie2017A-Submillimeter}, and the red solid line is their mean conversion, 
${\rm SFR_{IR}} (M_{\odot}~\rm{yr}^{-1})$$ = 143 \times S_{850\mu{\rm m}}$ (mJy). The red dashed lines correspond to a multiplicative factor of two about the solid red line. The median redshifts of the 
two samples are $z = 1.24$ (this work) and $z = 2.28$ (Cowie et al.). The large difference of conversion factors between the two studies is caused by the different SEDs of the two samples.}
\end{center}
\label{fig:figure6}
\end{figure}

\subsection{Individual Sources}

Here we describe some details for several galaxies that have special properties.

\subsubsection{0416-1 and 0416-3}

0416-1 is classified as an AGN in the Grism Lens-Amplified Survey from Space (GLASS; \citealt{Schmidt2014Through-the-Loo,Treu2015The-Grism-Lens-}). Because it is 
unclear whether the radio and UV emission is dominated by the AGN or star formation, we caution that the radio and UV SFRs for this source can only be considered as upper limits. 
The optical morphology of 0416-1 shows two peaks with a $\sim$ 0$\farcs$4 offset, suggesting that it might be a merger. 0416-3 has a pair of radio counterparts that are slightly blended at 3 GHz but clearly separated 
at 6 GHz. These two radio sources correspond to two galaxies that have photometric redshifts of $0.99\pm0.10$ and $1.01\pm0.10$. Both of the radio centers have $\sim$ 0$\farcs$5 offsets 
from the optical centers. These offsets and the consistent photometric redshifts suggest that the two galaxies are an interacting pair.

\subsubsection{0416-4 and 0717-2}

0416-4 and 0717-2 are the two sources without photometric redshifts. We show the $K_s$-band images of these two SMGs in Figure~7. Both of these sources are 
outside the WFC3 coverage and not detected by {\it Herschel}. The {\it HST} magnitudes of 0416-4 are only well measured in F425W, F606W, F814W, and F850LP; a photometric redshift of $z=$ 1.24$^{+1.99}_{-0.46}$ is reported in the CLASH catalog. The large difference between the observed brightness at optical wavelengths and $K_s$ band for this source suggests it is likely at high redshift. We obtained a redshift estimate of $z=$ 2.7$^{+2.7}_{-2.2}$ based on the 450 $\mu$m-to-850 $\mu$m flux ratio. 0717-2 is completely undetected in the {\it HST} images and also very faint in the Keck/MOSFIRE $K_s$-band image. This source would be an example of faint SMGs that are not included in the UV star formation history. Our redshift estimate for this source based on the 450 $\mu$m-to-850 $\mu$m flux ratio is $z=$ 4.5$^{+2.3}_{-1.3}$.

\begin{figure}
\begin{center}    
\includegraphics[width=4.2cm]{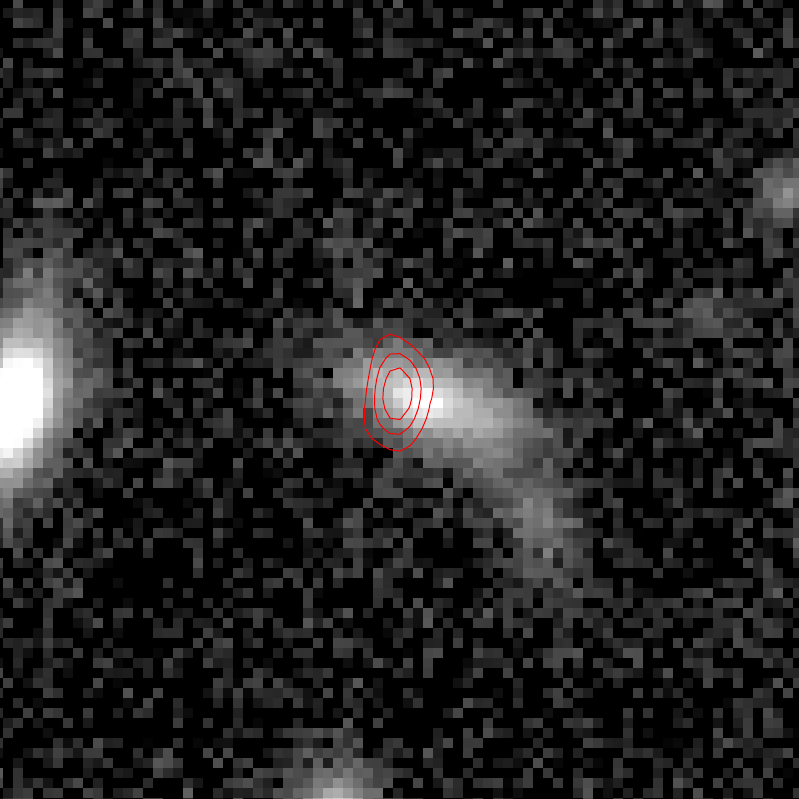}
\includegraphics[width=4.2cm]{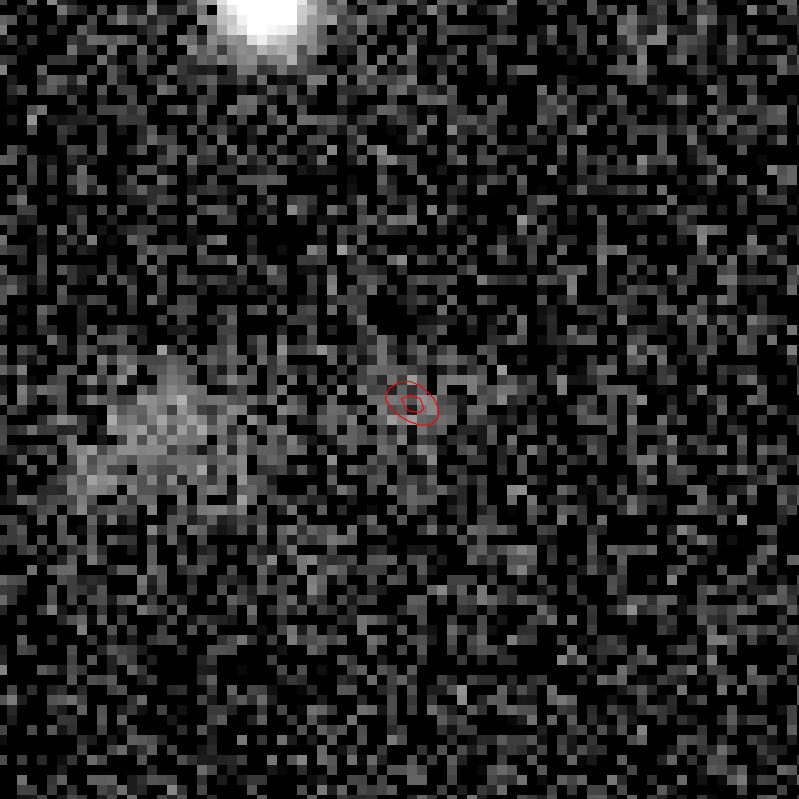}
\caption{$K_s$-band images \citep{Brammer2016Ultra-deep-K-S-} of 0416-4 (left) and 0717-2 (right) centered at the 3 GHz positions. The red contours are 
(3.0, 6.0, 9.0)$\times \sigma$ (left) and (3.0, 5.0)$\times \sigma$ (right) isophotes of the 3 GHz sources, where $\sigma \sim$ 1.0 $\mu$Jy~beam$^{-1}$ for both sources. The 
image size is 8$'' \times$ 8$''$}
\end{center}
\label{fig:figure7}
\end{figure}

\subsubsection{0717-1}

The radio position of 0717-1 is $\sim$ 1$\farcs$5 east from the center of a spiral galaxy at $z = 1.14$. A close-up {\it HST}/ACS image for this source in shown in Figure~8. This source is not within the 
WFC3 coverage. We can see faint and red structures at the radio position. It is not clear whether these structures are from a background lensed galaxy or are related to the spiral galaxy. We assumed the 
case of being related to the spiral galaxy to derive the properties of this SMG. In this case, the red structures might be the core and tidal tails of a smaller disrupted galaxy that is being merged into the 
larger spiral galaxy. Note that if this SMG is actually a background source at a higher redshift, the derived SFRs and dust temperature would all be higher.

\begin{figure}
\begin{center}    
\includegraphics[width=7cm]{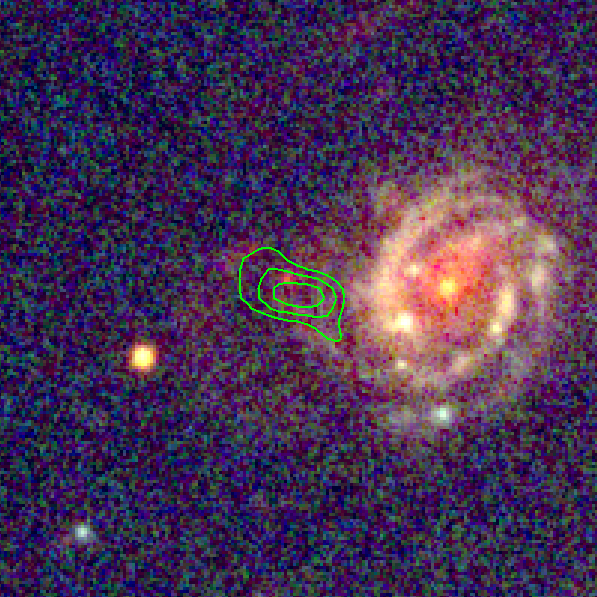}
\caption{Close-up {\it HST}/ACS false-color (F435W, F606W and F814W) image of 0717-1 centered at the 3 GHz position. The green contours are (3.0, 4.5, 6.0)$\times \sigma$ isophotes 
of the 3 GHz source, where $\sigma \sim$ 0.93 $\mu$Jy~beam$^{-1}$. The image size is 6$'' \times$ 6$''$}
\end{center}
\label{fig:figure8}
\end{figure}

\begin{figure}
\begin{center}    
\includegraphics[width=8cm]{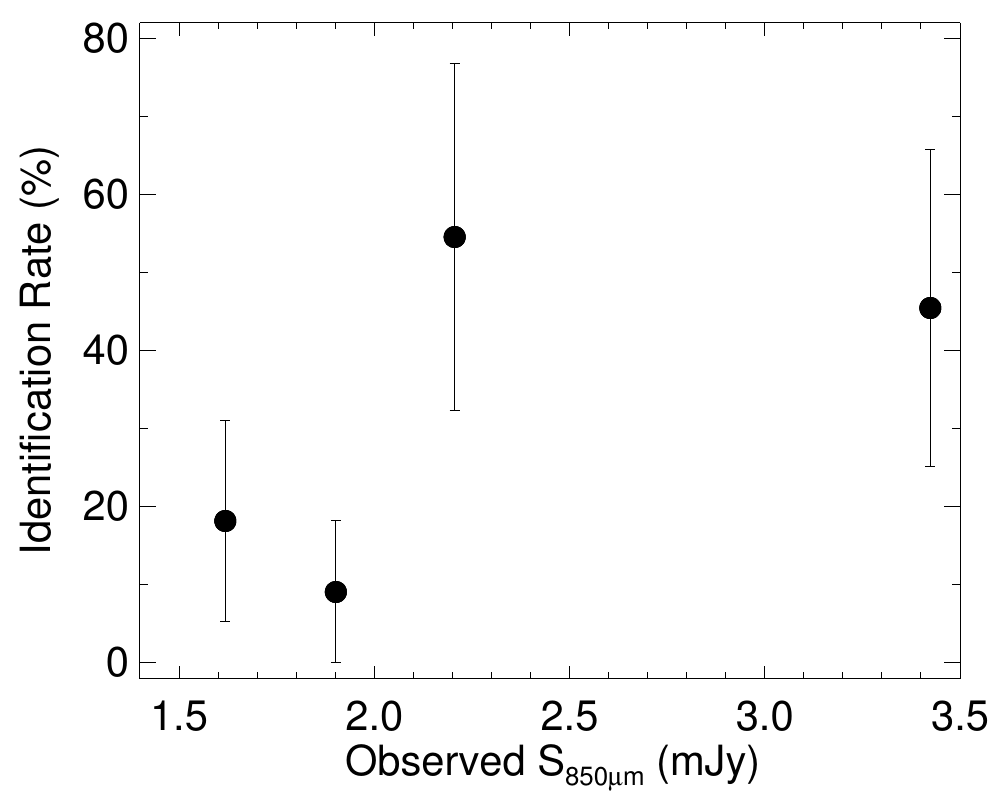}
\caption{Counterpart identification rate as a function of observed (not de-lensed) 850 $\mu$m flux density for the 44 SCUBA-2 sources within the {\it HST} coverage. The flux range is $\sim 1.4-5.7$ mJy. 
Each bin includes 11 sources and is plotted at the mean flux density. The errors are based on Poisson statistics.}
\end{center}
\label{fig:figure9}
\end{figure}  

\section{Detectability of the Submillimeter sources}

\subsection{Bias and Redshift Distribution}\label{sec:bias}

In the three Frontier Fields, there are 44 SCUBA-2 850 $\mu$m sources within the {\it HST} coverage, and we only found 15 radio counterparts to 14 of them. 13 of these 14 identified sources are detected in the optical images. All of the sources are detected in $K_s$ band, 3.6 $\mu$m, and 4.5 $\mu$m. 11 are at $z<2$, and the median redshift of the entire sample is $z = 1.28^{+0.07}_{-0.09}$ (0416-3 is counted as one source at $z=1.00$). This is much lower than the redshift distribution of the classical SMGs, which are typically found to be at $z = 2-3$ (e.g., \citealt{2011MNRAS.415.1479W,Smolcic2012Millimeter-imag,Casey2013Characterisatio,Weis2013ALMA-Redshifts-,Simpson2014An-ALMA-Survey-,Koprowski2015The-SCUBA-2-Cos}). The redshift distribution of our sample, which is lower than the classical 
SMGs, and the fact that we still miss about two-thirds of the SMGs in our radio images are caused by the bias of the radio identification technique. However, we note that some studies have suggested a ``cosmic downsizing'' \citep{Cowie1996New-Insight-on-} of SMG luminosities (e.g., \citealt{Heavens2004The-star-format,Bundy2006The-Mass-Assemb,Franceschini2006Cosmic-evolutio,Dye2008The-SCUBA-HAlf-,Mobasher2009Relation-Betwee,Magliocchetti2011The-PEP-survey:,Hsu2016The-Hawaii-SCUB,Cowie2017A-Submillimeter}). Therefore, fainter SMGs might indeed have a lower redshift distribution.

We show the counterpart identification rate as a function of observed 850 $\mu$m flux density in Figure~9. As expected, the identification rate is lower at fainter flux bins. Also, at the brightest flux 
bin ($\sim$ 3.4 mJy), 55\% of the sources are still not identified. Those sources without radio counterparts are likely at higher redshifts. To estimate the depth of our 3 GHz survey in terms of detecting faint SMGs, we consider a source with a specific observed 850 $\mu$m flux density. Assuming that the UV SFR is negligible and $\rm{SFR_{radio} \sim SFR_{IR}}$, we can use equation~(1) and our median $S_{850\mu{\rm m}}$--SFR$_{\rm IR}$ conversion in Section~\ref{sec:850SFR} to obtain the radio power $L_{\nu}$ at any rest-frame frequency. We can then compute the observed-frame 3 GHz flux density of this source as a function of redshift, as shown in Figure~10. This shows that, with our 5$\,\sigma$ detection limit of $\sim$ 5 $\mu$Jy beam$^{-1}$ at 3 GHz, we can only detect sources with observed $S_{850 \mu{\rm m}} =$ 2 mJy out to $z \sim 1.9$. A higher ${\rm SFR}_{\rm{IR}} / S_{850\mu{\rm m}}$ would lead to a higher redshift limit, which should be the case for the three sources at $z > 2$, 0416-1, 0416-4, and 0717-2. The value of ${\rm SFR}_{\rm{IR}} / S_{850\mu{\rm m}}$ for 0416-1 is 106. For 0416-4 and 0717-2, ${\rm SFR}_{\rm{IR}} / S_{850\mu{\rm m}}$ would be $>$ 200 if their IR SFRs agree with their radio SFRs.

We can also estimate a lower limit of the median redshift of all the 44 SCUBA-2 sources. Assuming all of the other 30 radio-faint SMGs are not blended multiples and they all have ${\rm SFR}_{\rm{IR}} / S_{850\mu{\rm m}} = 54$, we can compute the lower redshift limit for each of these sources to be detected by our 3 GHz images. Along with the 14 SMGs we already identified, the median redshift of the entire sample is at $z > 1.9$. In reality, some of these 30 sources would split into multiples, making 
the median redshift lower. Since a lower limit rather than an upper limit of the median redshift is estimated, it is not clear whether the redshift distribution of these cluster-lensed faint SMGs is indeed lower than those of the brighter samples. In addition, our 
estimated median redshift depends on the value of ${\rm SFR}_{\rm{IR}} / S_{850\mu{\rm m}}$. Because ${\rm SFR}_{\rm{IR}} / S_{850\mu{\rm m}}$ correlates with dust temperature (peak wavelength), the detectability of a SMG at 3 GHz is determined by both the dust temperature and the redshift. Therefore, future submillimeter interferometry is required to identify the multi-wavelength counterparts to the SCUBA-2 sources without radio counterparts, breaking the degeneracy of redshift and dust temperature distributions.

\begin{figure}
\begin{center}    
\includegraphics[width=9cm]{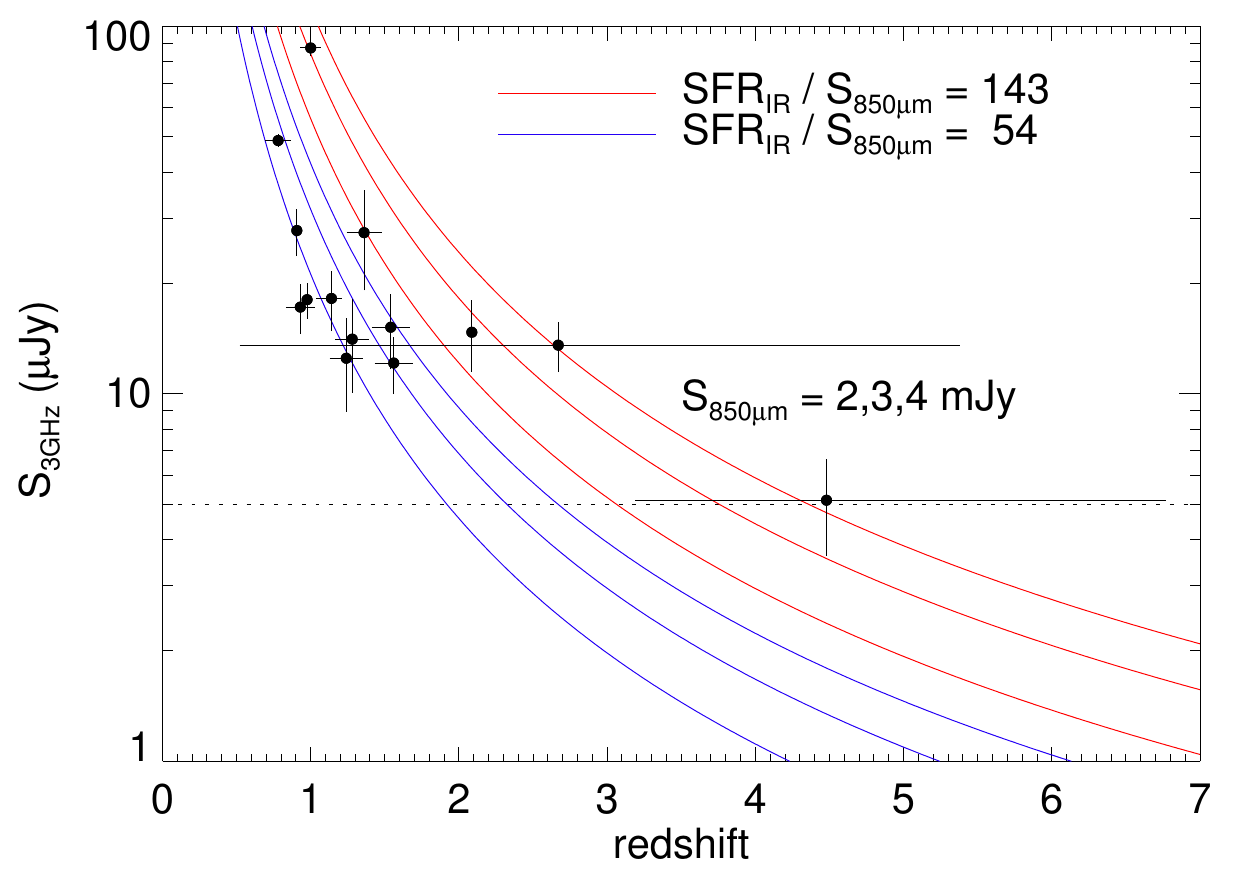}
\caption{Expected observed (not de-lensed) 3 GHz flux density as a function of redshift with different observed (not de-lensed) 850 $\mu$m flux densities 
based on the $S_{850\mu{\rm m}}$--SFR$_{\rm IR}$ conversions of this work (blue) and Cowie et al. (red; 2017). For each conversion, we plot 
the expected relations for $S_{850\mu{\rm m}} = $ 2, 3, and 4 mJy (bottom to top). The horizontal dashed line corresponds to our detection limit of $\sim$ 5 $\mu$Jy. 
Our sample of 14 SMGs are overplotted as black circles. We can only detect sources with observed $S_{850 \mu{\rm m}} =$ 2 mJy out to $z \sim 1.9$ if ${\rm SFR}_{\rm{IR}} / S_{850\mu{\rm m}} = 54$.}
\end{center}
\label{fig:figure10}
\end{figure}

\begin{figure}
\begin{center}    
\includegraphics[width=9cm]{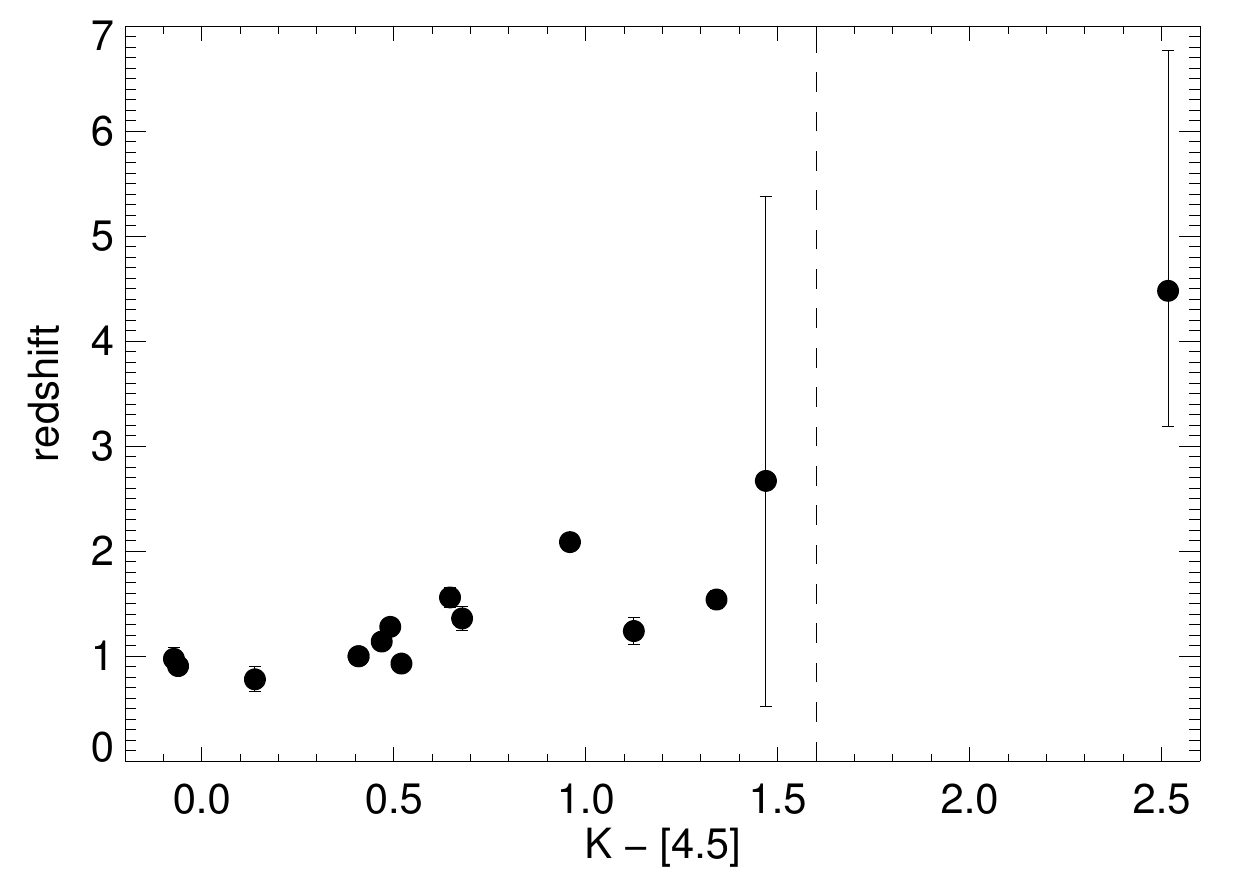}
\caption{$K$ -- [4.5] color \citep{Wang2012A-KS-and-IRAC-S} versus redshift for the 14 radio-identified SMGs. 0717-2 (the data point on the right) is the only source 
among our 14 sources that can be selected by $K$ -- [4.5] $>$ 1.6 (dashed line).}

\end{center}
\label{fig:figure11}
\end{figure}

\subsection{Optical-near-infrared Colors}

Several previous studies have shown that optical-near-infrared colors such as $i - K$, $J - K$, and $K -$[4.5] can effectively select high-redshift, dusty galaxies (e.g., 
\citealt{Smail2002The-nature-of-f,Dannerbauer2004The-Faint-Count,Frayer2004Near-Infrared-C,Caputi2012The-Nature-of-E,Wang2012A-KS-and-IRAC-S}). \cite{Cowie2017A-Submillimeter} 
showed that among their 22 radio sources that are selected by $K -$[4.5] $>$ 1.6 (KIERO; \citealt{Wang2012A-KS-and-IRAC-S}), 20 have submillimeter detections at the $>$ 3$\,\sigma$ 
level. \cite{Chen2016The-SCUBA-2-Cos} proposed a triple color cut (OIRTC) of $z-K$ $>$ 1.1 and $K-$[3.6] $>$ 1.25 and [3.6] $-$ [4.5] $>$ 0.22, which successfully selects sources from their 
ALMA training sample with an accuracy of 87\% and a completeness of 52\%.

We test both the KIERO and OIRTC techniques on our sample of radio-identified SMGs. Interestingly, only 0717-2, the source without an optical counterpart, can be selected by these 
two methods. In Figure~11, we can see a correlation between $K -$[4.5] color and redshift. Similar trends exist for $z-K$, $K-$[3.6] or [3.6] $-$ [4.5]
as well. This suggests that both of these color cuts pick out high-redshift red galaxies. As a result, they miss the galaxies in our low-redshift sample.

\section{Summary}

In this second paper of the Hawaii-S2LCS series, we cross-match our deep SCUBA-2 survey with VLA 3 and 6 GHz images for three {\it HST} Frontier Fields, MACS\,J0416.1--2403, MACS\,J0717.5+3745, and MACS\,J1149.5+2223. Within the {\it HST} coverage, 14 out of 44 SCUBA-2 850 $\mu$m sources have 5$\,\sigma$ detected 3 GHz counterparts. A close pair of radio counterparts are identified in one of the SCUBA-2 sources, so a total of 15 radio sources are detected. Only five of the SCUBA-2 sources (six of the radio sources) are detected at 6 GHz above a 5$\,\sigma$ level. The 850 $\mu$m flux densities of these sources span from 0.7 to 4.4 mJy after correcting for lensing amplification. We measure the dust temperatures, IR luminosities, and IR SFRs with our SCUBA-2 450 and 850 $\mu$m flux densities, the {\it Herschel} flux densities from \cite{Rawle2016A-complete-cens}, and the ALMA measurements at 1.1 mm from \cite{Gonzalez-Lopez2017The-ALMA-Fronti}. Radio and extinction-uncorrected UV SFRs are also computed based on our VLA imaging and the optical SEDs measured from the {\it HST} images. The radio SFRs well agree with the UV+IR SFRs.

These 14 faint SMGs are quite different from the classical, bright SMGs. First of all, the median redshift of our sample is $z = 1.28^{+0.07}_{-0.09}$, which is much lower than the typical values ($z=2-3$) in the 
literature. 13 out of the 14 sources would not be selected from the optical-near-infrared colors techniques KIERO \citep{Wang2012A-KS-and-IRAC-S} and OIRTC \citep{Chen2016The-SCUBA-2-Cos} due 
to their low redshifts. Secondly, we find that our sample has lower dust temperatures (longer $\lambda_{\rm peak}$) than those of the bright SMGs. This is also confirmed by the lower values of ${\rm SFR}_{\rm{IR}} / S_{850\mu{\rm m}}$. However, these 14 sources may not represent the general submillimeter population at the same flux range, given that the SCUBA-2 sources without radio counterparts are 
likely at higher redshifts. Future submillimeter interferometry is required to identify the multi-wavelength counterparts to these radio-faint sources, creating an unbiased sample of faint SMGs for 
more statistical studies.

 \acknowledgments

We gratefully acknowledge support from NSF grants AST-0709356 (L.-Y. H. and L. L. C.) and AST-1313150 (A. J. B.), as well as the John Simon Guggenheim Memorial Foundation and Trustees of 
the William F. Vilas Estate (A. J. B.). I. R. S. acknowledges support from STFC (ST/L00075X/1), the ERC Advanced Investigator programme DUSTYGAL (321334) and a Royal Society/Wolfson Merit Award. The 
James Clerk Maxwell Telescope is operated by the East Asian Observatory on behalf of The National Astronomical Observatory of Japan, Academia Sinica Institute of Astronomy and Astrophysics, the Korea Astronomy and Space Science Institute, the National Astronomical Observatories of China and the Chinese Academy of Sciences (Grant No.\,XDB09000000), with additional funding support from the Science and Technology Facilities Council of the United Kingdom and participating universities in the United Kingdom and Canada. The James Clerk Maxwell Telescope has historically been operated by the Joint Astronomy Centre on behalf of the Science and Technology Facilities Council of the United Kingdom, the National 
Research Council of Canada and the Netherlands Organisation for Scientific Research. Additional funds for the construction of SCUBA-2 were provided by the Canada Foundation for Innovation. The National 
Radio Astronomy Observatory is a facility of the National Science Foundation operated under cooperative agreement by Associated Universities, Inc. We acknowledge the cultural significance that the summit 
of Mauna Kea has to the indigenous Hawaiian community.

\bibliographystyle{apj}
\bibliography{ms}

\end{document}